\documentclass[aps,superscriptaddress,abstract,nofootinbib]{revtex4}
\usepackage{graphics}
\usepackage{amsmath}
\graphicspath{{Figures/},{Logos/}}
\usepackage{graphicx,color}

\DeclareMathOperator{\csch}{csch}
\usepackage{bm}
\usepackage{amsfonts}
\usepackage{amsmath}
\usepackage{amssymb}
\usepackage{float}
\usepackage{hyperref}
\usepackage{subfigure}
\usepackage{slashed}

\begin{document}

\title{Parity violating gravitational waves at the end of inflation}

\author{Mar Bastero-Gil} \email{mbg@ugr.es} \affiliation{Departamento
  de F\'{\i}sica Te\'orica y del Cosmos, Universidad de Granada,
  Granada-18071, Spain}

\author{Ant\'onio Torres Manso} \email{atmanso@correo.ugr.es} \affiliation{Departamento
  de F\'{\i}sica Te\'orica y del Cosmos, Universidad de Granada,
  Granada-18071, Spain}

\begin{abstract}
Inflaton-vector interactions of the type $\phi F\tilde{F}$ have provided interesting phenomenology to tackle some of current problems in cosmology, namely the vectors could constitute the dark matter component.
It could also lead to  possible signatures imprinted in a gravitational wave spectrum.  Through this coupling, a rolling inflaton induces an exponential production of the transverse polarizations of the vector field, having a maximum at the end of inflation when the inflaton field velocity is at its maximum.
These gauge particles, already parity asymmetric, will source the tensor components of the metric perturbations, leading to the production of parity violating gravitational waves. 
In this work we examine the vector particle production in the weak coupling regime, integrating the gauge mode amplitudes spectrum during the entirety of its production and amplification epochs, until the onset of radiation domination. Finally, we calculate the gravitational wave spectrum combining the vector mode analytical solution, the WKB expansion, valid only during the amplification until horizon crossing, and the numerical solution obtained at the beginning of radiation domination when the modes cease to grow.
\end{abstract}
\maketitle

\section{Introduction}
Cosmological inflation, an early phase of accelerated expansion,  is currently the preferred solution to address the flatness and horizon problem in Standard Cosmology \cite{infGuth,inflinde,infalbrecht}. Typically a scalar field, the inflaton $\phi$, during a slow-roll phase drives such an expansion, and through its quantum vacuum fluctuations gives a natural mechanism to generate the observed anisotropies in the Cosmic Microwave Background (CMB). The question on how to move into the better known Standard Cosmology has to be addressed, including the period known as reheating, the transition into a radiation dominated universe, where  production of light nuclei at Big Bang Nucleosynthesis (BBN) takes place \cite{Fields:2014uja,ParticleDataGroup:2020ssz}. To do so one has to couple the inflaton with other particles species \cite{Figueroa:2018twl}.

Inspired by "axion-like" inflation models \cite{Pajer:2013fsa}, where a shift symmetry protects the flatness of the inflaton potential, couplings with $U(1)$ vector particles $\alpha \phi F \tilde{F}/f$ are often considered, where $\alpha$ is the coupling constant and $f$ an energy scale. As $\phi$ rolls down in the slow-roll evolution it will source a tachyonic amplification of the vector modes from their vacuum fluctuations into a classical state. 
As the interaction parameter will depend linearly on the inflaton velocity, the largest amplification is expected at the end of inflation, as the system escapes the slow-roll evolution\footnote{In regimes with large interaction parameters and considering the backreaction of vector production on inflaton evolution, there are non-linear effects which affects the dynamics and the analysis is not as straightforward, see \cite{Cheng:2015oqa}.}. 
Due to the parity violating nature of the interaction only one of the vector transverse degrees of freedom is amplified \cite{Anber:2006xt}. This abrupt production of gauge fields may source a sizable production of gravitational waves (GW) \cite{Sorbo:2011rz,Anber:2012du,Garcia-Bellido:2016dkw,Bartolo:2016ami}, also parity asymmetric, within a range of frequencies that will depend on the stage of inflation\footnote{Parity violating GWs have also been proposed in the context of a gravitational Chern Simons term, as discussed in \cite{Lue:1998mq,Alexander:2004wk,Contaldi:2008yz,Cai:2016ihp,Odintsov:2022hxu}}. This could mean an observational signal in the CMB or at interferometer scales.  

Phenomenology with these models is extremely broad. For example the production of non-Gaussianities in the primordial comoving curvature perturbations has been studied in \cite{Barnaby:2010vf,Barnaby:2011vw,Linde:2012bt}, and provides the strongest bound on the interaction parameters at CMB scales. The inflaton-vector dynamics has also been used to study the production of primordial magnetic fields \cite{Field:1998hi,Anber:2006xt,Adshead:2016iae}, as a dissipation mechanism allowing inflation with steeper potentials \cite{Anber:2009ua}, or as an ignition for warm inflation in the case of Yang-Mills gauge interactions \cite{mwi,laine}. Other studies focus on the generation of different dark sectors, like primordial black hole production \cite{Linde:2012bt,Garcia-Bellido:2016dkw} but also particle dark sectors \cite{Nelson:2011sf,Arias:2012az,vectorDM,vectorDM1,Nakayama:2020rka,Nakai:2020cfw,Salehian:2020asa,Firouzjahi:2020whk}. Within the later, we may  have the vectors as a dark matter candidate, compatible with the observed dark matter relic abundance for masses  $\mu \mathrm{eV} \lesssim m \lesssim 10\, \mathrm{TeV}$ \cite{vectorDMown,Bastero-Gil:2021wsf},  or due to the extremely efficient pair production via Schwinger effect when including a fermion-vector coupling \cite{Domcke:2018eki,Domcke:2019qmm}. The parity asymmetry within the system has also been exploited to explain the baryon asymmetry in the Universe \cite{Giovannini:1997eg,Anber:2015yca,Fujita:2016igl,Cado:2016kdp,Jimenez:2017cdr,Domcke:2019mnd}.

In this work we will examine the gravitational wave spectrum sourced by the inflaton-vector coupling not only during inflation, but until the onset of a radiation dominated era. End of inflation (accelerated expansion) is set by the condition $\epsilon_H=-\dot H/H^2=1$, $H$ being the Hubble expansion rate and $\dot H$ its time derivative, whereas in a radiation dominated universe one has $\epsilon_H=2$. In this extra period from $\epsilon_H=1$ to $\epsilon_H=2$, one  still finds amplitude enhancement in the larger momentum modes that are still sub-horizon at the end of inflation. Indeed, vector particle production can easily take place during preheating \cite{Adshead:2015pva,vectorDM1, Co:2018lka}, i.e., the first stages of the reheating period \cite{preheating, preheating2}, but typically for coupling values $\alpha m_P/f$ larger than those required to avoid backreaction (BR) effects during inflation. Gravitational wave production has been extensively studied in this preheating regime \cite{Adshead:2018doq,Adshead:2019igv,Adshead:2019lbr,Weiner:2020sxn}.  We will therefore stay within the linear, non-backreaction (NBR) regime, for which preheating effects can be ignored. Nevertheless, our main point is that even in this regime where we may expect to be able to treat the transition to radiation perturbatively, particle production continues up to $\epsilon_H=2$ invalidating the linear analyses. Taking into account this regime, we aim to derive an upper bound on $\alpha m_P/f$ for which non-linear effects may be ignored. 

We will take a semi-analytical approach to describe the gauge mode amplitudes that result from the tachyonic amplification sourced by $\phi$. It will consist on combining the analytical solution valid to describe the vector amplitudes during the amplification until horizon crossing and the solution obtained numerically at the beginning of radiation domination, when the modes cease to grow. In our analysis we will see that in order to avoid backreaction effects due to vector production at the latter stages of inflation, one must take an interaction parameter $\xi=(\alpha/f)( \dot \phi/H)/2$ at 60 e-folds before the end of inflation smaller than what is constrained by non-Gaussianities ($\xi_{60}\lesssim 2.5$). We will try a simple scheme, based on energy conservation of the vector modes, to mimic the backreaction effects on the inflaton motion without doing the individual numerical integration for every vector. This will allow us to study systems with  $\alpha\, m_P/f \lesssim 16$ ($\xi_{60}\lesssim 0.16$) for the $\alpha$-attractor model with a good description.  We will find a spectrum of gravitational waves with a peak at very large frequencies,  $10^7~ {\rm Hz}\lesssim f\lesssim 10^9~ {\rm Hz}$, typical of production mechanisms at the end of inflation and (p)reheating.

There has been an effort to study the full non-linear regime with different methodologies. Through what is called the gradient expansion formalism, a truncated system of bilinear functions of the electric and magnetic fields in position space, one mimics the backreaction effects and can source the dissipation on the inflaton dynamics. In \cite{Sobol:2020lec,Gorbar:2021rlt} these integrations were performed until the end of inflation at $\epsilon_H=1$, with a quadratic inflaton potential, thus they do not account for the effects during (p)reheating until $\epsilon_H=2$. Nevertheless, interesting dynamics results from the backreaction estimation, namely on the oscillating effect on the interaction parameter $\xi$ at the end of inflation and the double peak structure at the spectral energy density of the vector modes. In the works \cite{Cheng:2015oqa,DallAgata:2019yrr,Domcke:2020zez,Peloso:2022ovc,Garcia-Bellido:2023ser} the non-linear evolution of $\xi$ was also obtained, and recently confirmed in \cite{Caravano:2022epk, Caravano:2022yyv,Figueroa:2023oxc} after the first lattice computations on gauge particle production in axion inflation.

This paper is organized as follows. In section \ref{secII} we set the model and  study the vector production to discuss the validity of the non-backreaction description.  We provide a zero order attempt to manage the backreaction effects in section \ref{secIII}. Finally, we calculate the gravitational wave spectrum in section \ref{secIV}, to then conclude and discuss followup work in section \ref{secV}. Technical details about the parametrization used for the vector power spectrum and the calculation of the induced GW spectrum are given respectively in Appendix \ref{appendixA} and \ref{appendixB}. We also provide a comparison between the results for an $\alpha$-attractor inflationary model and the standard evolution with a quartic potential in \ref{appendixC}

\section{Vector particle production}\label{secII}

Consider a system described by the action
\begin{equation}
{\cal{S}}=-\int d^{4} x \sqrt{-g}\left[\frac{1}{2} \partial_{\mu} \phi \partial^{\mu} \phi+V(\phi)+\frac{1}{4} F_{\mu \nu} F^{\mu \nu}+\frac{1}{2} m^{2} A_{\mu} A^{\mu}+\frac{\alpha}{4 f} \phi F_{\mu \nu} \tilde{F}^{\mu \nu}\right]
\end{equation}
where the potential $V(\phi)$ drives the slow-roll evolution, $\alpha/f$ quantifies the inflaton-vector coupling, $F_{\mu\nu}$ is the field strength and $\tilde{F}^{\mu \nu} $ its dual, $\tilde{F}^{\mu \nu}=\epsilon^{\mu \nu \alpha \beta} F_{\alpha \beta} / (2\sqrt{-g})$. We use the Friedmann-Robertson-Walker metric with ${ds^2=-dt^2+a^2(t)dx^2}$ and the convention  $\epsilon^{0123}=1/\sqrt{-g}$. The vector mass can be of a Stueckelberg type or be produced through a symmetry breaking phase transition. It will be considered to be smaller than the Hubble scale at the end of inflation, making it negligible during the tachyonic production.

In our analysis we will consider an $\alpha$-attractors potential $V(\phi)=(9\lambda/4)\tanh^4[\phi/(\sqrt{6}\, m_P)]m_P^4$ \cite{Kallosh:2013yoa}, allowed by Planck data \cite{planckinflation}, which at the end of inflation and reheating will tend towards the quartic potential $V(\phi)=\lambda\phi^4/4$ \cite{Bastero-Gil:2020uww}. 

In order to study the production of gauge particles induced by the rolling inflaton, we promote the classical field $A(t,x)$ to an operator $\hat{A}(t,x)$, to then be expanded in terms of creation and annihilation operators and the mode functions in an helicity basis
\begin{equation}
  \hat{A}_{i}(t, \mathbf{x})=\int \frac{d^{3} \mathbf{k}}{(2 \pi)^{3 }} e^{i \mathbf{k} \cdot \mathbf{x}} \hat{A}_{i}(t, \mathbf{k})=\sum_{\lambda=\pm,L} \int \frac{d^{3} \mathbf{k}}{(2 \pi)^{3 }}\left[\epsilon_{\lambda}^{i}(\mathbf{k}) A_{\lambda}(t, \mathbf{k}) \hat{a}_{\lambda}^{\mathbf{k}} e^{i \mathbf{k} \cdot \mathbf{x}}+\mathrm{h.c.}\right]
\,,
  \label{Eq:Gauge_expansion}
\end{equation}
where we  have separated the three degrees of freedom of the vector into transverse and longitudinal components $\bar{A}_T$ and $A_L$ respectively, $\bar{k}\cdot\bar{A}=kA_L$
and $\bar{k}\cdot\bar{A}_T=0$. Furthermore, we have written the transverse component in terms of the two helicities, $\bar{A}_{T}=\bar{\epsilon}_{+}A_+ +\bar{\epsilon}_{-}A_-$. The creation and annihilation operators satisfy the commutation relations,
\begin{equation}
\left[a_{\lambda}(\vec{k}), a_{\lambda}^{\dagger}\left(\vec{k}^{\prime}\right)\right]=(2 \pi)^{3} \delta_{\lambda \lambda^{\prime}} \delta^{3}\left(\vec{k}-\vec{k}^{\prime}\right) \,.
\end{equation}

The time components of $A$ for both longitudinal and transverse polarizations have been recently derived in Appendix A of \cite{Bastero-Gil:2021wsf} and are given by
\begin{flalign}
A_0^{\ L}(\vec{k}, \tau)  &=\frac{-i k \cdot \partial_\tau A_L(\vec{k}, \tau)}{k^2+a^2 m^2},  \\
A_0^{\ \pm}(\vec{k}, \tau)  &=0\,.
\end{flalign}
As a result, the time component of the vector field does not mix the transverse and longitudinal components.
The scalar field and the vector mode spatial components, in Fourier space, will follow the equations of motion \cite{Anber:2009ua,Barnaby:2010vf,Bastero-Gil:2021wsf}
\begin{flalign}
&\ddot{\phi}+3 H \dot{\phi}+V^{\prime}(\phi)=\frac{\alpha}{4 f}\langle F \tilde{F}\rangle\,, \\
&\ddot{A}_{\pm}+H \dot{A}_{\pm}+\left(\frac{k^{2}}{a^{2}} \pm \frac{k}{a} \frac{\alpha \dot{\phi}}{f}+ m^2\right) A_{\pm}=0\,, \\
&\ddot{A}_{L}+\frac{3k^2+a^2m^2}{k^2+a^2m^2}H \dot{A}_{L}+\left(\frac{k^{2}}{a^{2}}+m^2\right) A_{L}=0\,,
\end{flalign}
where the overdots denote derivatives with respect to physical time $t$ and $k\equiv|k|$ is the magnitude of the comoving momentum. We consider only the spatially homogeneous zero momentum mode $(k = 0)$ of the inflaton.

Immediately, one observes how the inflaton motion enters into the equations of motion for the transverse modes. We will take the vector mass to be smaller than the Hubble scale at the end of inflation and so that its effects are negligible for the tachyonic production, see \cite{Bastero-Gil:2021wsf}, and as a result having also little effect on the GW generation. As for the longitudinal mode it is not affected by the presence of the coupling with the inflaton, nonetheless it may be produced via inflationary fluctuations, as described in  \cite{vectorDM}. At the end of inflation  the energy density can be estimated as
\begin{equation}
    \frac{\rho_{A_L}}{\rho_\phi}=\frac{H_{end}^4}{4(2\pi)^2}\frac{1}{ \rho_\phi}\simeq 
    \frac{1}{24\pi^2}\left(\frac{H_{end}}{m_P}\right)^2
    \simeq6 \times10^{-18}  \left(\frac{\lambda}{10^{-14}} \right)  \,,
\end{equation}
where we took $H_{end}^2=V(\phi_{end})/(3m_P^2)$.
Thus, the longitudinal modes do not have a relevant contribution on the expansion and, as they do not mix with the transverse modes, will not play any role on the backreaction on the inflation motion.
Therefore, for the sake of simplicity, for the rest of the analysis we will neglect the effects of the vector mass and the evolution of the longitudinal mode during inflation, and for simplicity we set the mass to zero. We can then write the transverse modes equation of motion in  conformal time defined as $a d\tau=dt$,
\begin{equation}
  \left[\frac{\partial^{2}}{\partial \tau^{2}}+k^{2} \pm 2 k \frac{\xi}{\tau}\right] A_{\pm}( k,\tau) =0,\qquad\quad\xi \equiv \frac{\alpha \dot{\phi}}{2 H f}=\sqrt{\frac{\epsilon}{2}} \frac{\alpha}{f} m_{\mathrm{P}}\,.
  \label{Eq:A_eom_conformaltime}
\end{equation}
where  $\tau\simeq-1/(aH)$ during inflation, 
and $\epsilon \equiv-\dot{H} / H^{2}$, which for single field inflation is given by 
\begin{equation}
\epsilon\simeq\frac{\dot{\phi}^{2}}{2 H^{2} m_{P l}^{2}} \,.
\end{equation}
 Depending on the sign of the interaction parameter, $\xi$, one of the modes will experience tachyonic enhancement, when 
\begin{equation}
k^{2}\pm2 k \frac{\xi}{\tau}=k^{2}\mp2 k \xi a H<0 \,. 
\end{equation}
Using the convention $\dot{\phi}>0$, it results in $\xi>0$, implying that only the $A_+$ mode will develop an instability, while $A_-$ will stay in vacuum. 

Treating $\xi$ as constant, appropriate during a slow-roll evolution one can solve Eq. \eqref{Eq:A_eom_conformaltime} analytically in terms of Coulomb functions \cite{axioninf}. In the tachyonic regime which we are interested in, ${-k\tau< 2\xi\, (k < 2\xi aH)}$, and for $1/(8\xi)<-k\tau$ the Coulomb functions are  well approximated by the WKB expansion
\begin{equation}
  A_{+}(k, \tau)_{\mathrm{WKB}} \simeq \frac{1}{\sqrt{2 k}}\left(\frac{-k \tau}{2 \xi}\right)^{1 / 4} e^{\pi \xi-2 \sqrt{-2 \xi k \tau}} \,.
  \label{Eq:WKB_vector}
\end{equation}
This analytical expression provides a good intuition into the behavior of the modes around horizon crossing, when they experience the tachyonic enhancement \cite{Anber:2009ua,Barnaby:2011vw}. However, for regions outside $ 1/(8\xi)<-k\tau< 2\xi$, Eq. \eqref{Eq:WKB_vector} is not expected to provide a reliable description.

\begin{figure}[H]
	\centering
	\includegraphics[totalheight=6cm]{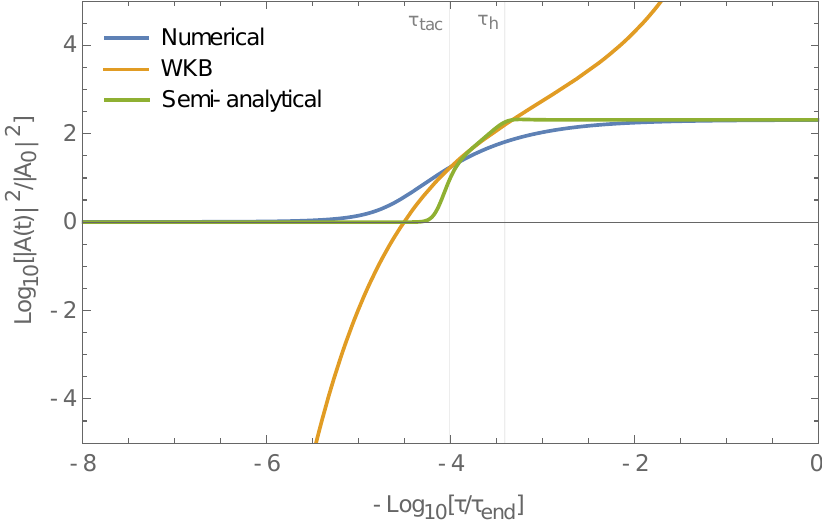}
	\caption{ Comparison of the time evolution, in a normalized conformal time, of the WKB expansion Eq. \eqref{Eq:WKB_vector}, with the numerical solution for  $|A_k|^2$ and the semi-analytical approximation Eq. \eqref{Eq:stepfunction} for a fixed mode with $k\simeq4\times 10^{-5} (aH)_{end}$ and $\alpha\, m_P/f=20$}
	\label{fig:WKBvsNumerical}	
\end{figure}

We compare the evolution of the approximation in Eq. \eqref{Eq:WKB_vector} with the numerical solutions of the system of equations in Fig. \ref{fig:WKBvsNumerical}. As expected we see that the WKB approximation describes well the time evolution during the tachyonic growth, but fails to describe the numerical results both when the modes go out of the horizon, and the initial vacuum state. In regards to the numerical integration one obtains what is theoretically predicted, a three phase function: first in the vacuum state, then the tachyonic growth at $\tau_{tac}=-\xi/k$, to then almost freezing at the horizon crossing after $\tau_h\simeq-1/(10 k)$.
 
With the aim of correctly describing the behavior of the gauge modes amplitudes and velocities, we build a semi-analytical solution. We combine the known analytical expressions in the vacuum state and the WKB solution during the tachyonic enhancement, with the amplitudes of $A$ and $A'$ at the end of inflation, for which we will use the subscript \textit{"end"}
Thus, from an arbitrary $\tau$ to $\tau_{end}$, we use a step function that shall give 
\begin{equation}
A_+(k,\tau)=
 \begin{cases}
{A_{BD}(k)} & \tau<\tau_{tac}\\
{A_{WKB}(k,\tau)} & \tau_{tac}<\tau<\tau_h\\
A_+(k,\tau_{end}) & \tau> \tau_h\ \,,
\end{cases}\label{Eq:stepfunction}
\end{equation}
with an analogous function for $A'_+(k,\tau)$. Finally, to obtain $A_+(k,\tau_{end})$ one must integrate the full numerical system for several modes $k$ to obtain the spectrum at $\tau_{end}$. We compare in Fig. \ref{fig:WKBvsNumerical} the semi-analytical approximation with the WKB and the numerical descriptions.  



Recalling the inflaton equation of motion, on the right-hand side one has the backreaction of the gauge modes on the inflaton evolution
\begin{flalign}
&\ddot{\phi}+3 H \dot{\phi}+V^{\prime}(\phi)=\frac{\alpha}{4 f}\langle F \tilde{F}\rangle\,.
\end{flalign}
Typically backreaction effects are  neglected if $\xi< {\cal{O}}(10)$
\cite{axioninf,Bastero-Gil:2021wsf}, with $\xi$ defined in Eq. \eqref{Eq:A_eom_conformaltime}. Moreover, $\xi$ is constrained from the non-observation of non-gaussianities in CMB measurements to be $\lesssim2.5$ \cite{Barnaby:2011vw,Barnaby:2010vf} at these scales (around 60 e-folds from the end of inflation). Naturally, this will constrain the  coupling constant $\alpha m_{P}/f$. We then write $\xi_{60}$ as the value of $\xi$ at 60 e-folds before the end of inflation to fix such constants. In the case of slow-roll inflation with the $\alpha$-attractors potential and no-backreaction ones has  
\begin{equation}
	\xi_{60}=2\sqrt{\frac{2}{3}}\, \frac{\alpha m_{P}}{f} \csch\left(\sqrt{\frac{2}{3}}\frac{\phi_{60}}{m_P}\right)\,.\label{Eq:xiCMB}
\end{equation}

\begin{figure}[H]
\centering
\includegraphics[totalheight=5.6cm]{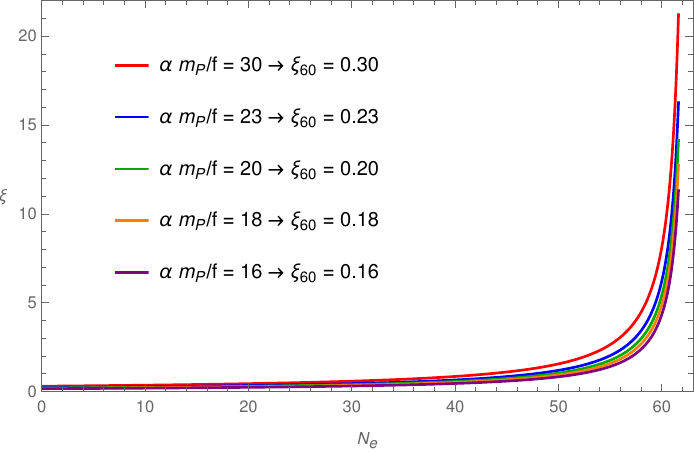} \includegraphics[totalheight=5.7cm]{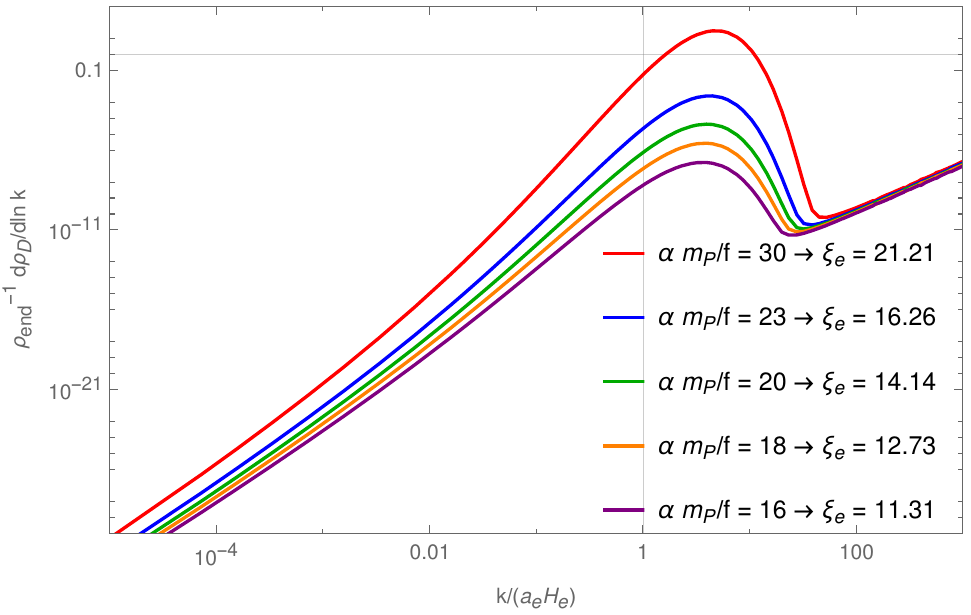}
\caption{Left panel: $\xi$ evolution under no backreaction effects,  we consider different values for the inflaton-vector coupling $\alpha/f$, fixing $\xi$ at 60 e-folds before the end of inflation as indicated in the plot. Right panel: Vector energy density spectrum normalized with the inflaton energy density at the end of inflation.
  The values for the interaction parameter at the end of inflation $\xi_e$ are also indicated for each $\alpha/f$. }
\label{fig:xievolutionnbr}	
\end{figure}

 Studying  $\xi$ evolution within the allowed parameter range one easily finds that from an initial non-backreacting regime the system evolves at the end of inflation to a relevant backreaction scenario, see Fig. \ref{fig:xievolutionnbr}. As expected from the proportionality with the inflaton velocity we get the maximum at the end of slow-roll inflation.
In order to test if the system reaches a backreaction regime before the end of inflation, we may calculate the energy density spectrum for the vector modes\footnote{Deriving the stress-energy tensor $T_{\mu\nu}$ from the action there are two terms proportional to $\phi F\tilde{F}$ which cancel exactly. As a result, the operator responsible for inducing the tachyonic
instability will not contribute to the total energy density. Nonetheless, as described in Section \ref{secIII}, such a term will be relevant and mediates the energy transfer between the inflaton and the vector fields. } $\rho_A$ (see Appendix A of \cite{Bastero-Gil:2021wsf})

\begin{flalign}
\rho_{A} &=\frac{1}{2}\langle\vec{E}^{2}+\vec{B}^{2}\rangle\nonumber \\
&=\frac{1}{4 \pi^{2} a^{4}} \int_{0}^{\infty} d k\, k^{2}\left(\left|\partial_{\tau} A_{+}(k, \tau)\right|^{2}+k^{2}\left|A_{+}(k, \tau)\right|^{2}\right)\nonumber \\ &=\frac{1}{2 a^{4}} \int d \ln k\left(\mathcal{P}_{\partial_{\tau} A_{+}}(k, \tau)+k^{2} \mathcal{P}_{A+}(k, \tau)\right)\,, \label{Eq:energydensityint}
\end{flalign}
where we have neglected the subdominant contributions from $A_-$, and defined the power spectrum 
\begin{equation}
\mathcal{P}_{X}(k, \tau)=\frac{k^{3}}{2 \pi^{2}}|X(k, \tau)|^{2} \quad X=A_{+} \text { or } \partial_{\tau} A_{+} \,, 
\end{equation}
with $\rho_{A}\equiv\langle\rho_{A}\rangle$ as the spatial average. Solving numerically for the equation of motion we can compute the quantity
\begin{equation}
\frac{1}{\rho_{end}}\frac{d \rho_{A}}{d \ln k}=\frac{1}{\rho_{end}}\frac{1}{2 a^{4}}\left(\mathcal{P}_{\partial_{\tau} A_{+}}(k, \tau)+k^{2} \mathcal{P}_{A_{+}}(k, \tau)\right)
\end{equation}
at the end of inflation, where $\rho_{end}$ corresponds to the total energy density in the universe. In a valid approximation, this ratio cannot be larger than one, i.e. the energy density of a single mode ought not to overcome the total energy density in the universe, here $\rho_{end} \simeq \rho_{\phi}$.
The vector energy density spectrum at the end of inflation, normalized by the inflaton energy density, is shown on the RHS in Fig. \ref{fig:xievolutionnbr}. One can then see that already values of $\xi_{60} \geq 0.30 $ will require taking into account backreaction effects to study gauge mode enhancement close to the end of inflation. 

Finally, as visible  in the right panel of Fig \ref{fig:xievolutionnbr} , the integral in  Eq. \eqref{Eq:energydensityint} will diverge. The divergence comes from the vacuum contribution as $\rho_{A_{BD}}\propto\Lambda^4$, with $\Lambda$ being a cut off scale, that can be addressed with a proper vacuum subtraction \cite{Parker:2009uva}. It can be seen from the same figure that the effects of the tachyonic amplification do not contain extra  divergent contributions as UV modes will not be enhanced. In practice one can use a hard cut off on the integration taking $\Lambda$ as the larger mode to be amplified. As discussed in section \ref{secIV}, when parametrizing the amplification we will introduce a regulator, a decaying exponential, to softly modulate this divergence. 

\section{Backreaction on the inflaton evolution at $0^{th}$- order}\label{secIII}

As seen in the previous section, from an initial non-backreacting dynamics at 60 e-folds, with the progression of slow-roll inflation the interaction parameter $\xi$ will tend to grow into a backreacting regime. 
The equations of motion together with the Bianchi identities in terms of the electric and magnetic fields in conformal time are given by 
\begin{flalign}
	&\phi''+2a H \phi' -\nabla^2\phi +a^2 \frac{d V(\phi)}{d\phi}=\frac{\alpha}{f}a^2\bar{E}\cdot\bar{B} \,, \\
	&\bar{E}'+2aH\bar{E}-\nabla \times \bar{B}=-\frac{\alpha}{f}\phi'\bar{B}-\frac{\alpha}{f}\bar{\nabla}\phi\times\bar{E} \,, \label{Eq:electricfield}\\
	&\bar{B}'+2aH\bar{B}+\bar{\nabla}\times\bar{E}=0 \,, \label{Eq:magneticfield}\\
	&\bar{\nabla}\cdot\bar{E}=-\frac{\alpha}{f}\bar{\nabla}\phi\cdot\bar{B}\\
	&\bar{\nabla}\cdot\bar{B}=0 \,, 
\end{flalign}
with 
\begin{flalign}
	&a^2 \bar{B}=\bar{\nabla}\times\bar{A} \,,\\
	&a^2 \bar{E}=-\bar{A}'+a\bar{\nabla}A_0 \,,
\end{flalign}
where $'$ denotes a derivative in conformal time $\tau$.

We now try a method to control the backreaction effects without relying on the  WKB time derivative expansion, describing the electric contribution from the gauge modes. In essence we will rely on energy conservation of the vector modes, 
 to measure its impact on the background dynamics. The energy density evolution is given by
\begin{flalign}
	&\dot{\rho}_\phi+3 H (\rho_\phi+p_\phi)=\dot{\phi}\frac{\alpha}{f}S_{EB}\label{Eq:eominflationenergydensity_BR} \,, \\
	&\dot{\rho}_A+3 H (\rho_A+p_A)=-\dot{\phi}\frac{\alpha}{f}S_{EB}\label{Eq:eomAenergydensity_BR}	\,,
\end{flalign}
where $S_{EB}=\left\langle\bar{E}\cdot\bar{B}\right\rangle$ is the source term. This interaction term will mediate the transfer of energy between the inflaton and the gauge fields. 
With Eqs. \eqref{Eq:electricfield} and \eqref{Eq:magneticfield} one can study its time evolution,
\begin{flalign}
	\dot{S}_{EB}=\dot{\bar{E}}\cdot\bar{B}+\bar{E}\cdot\dot{\bar{B}}=-4H\bar{E}\cdot\bar{B}-\frac{\alpha}{f}\dot{\phi}\left|\bar{B}\right|^2\,,
\end{flalign}
resulting in 
\begin{flalign}
\dot{S}_{EB}+4HS_{EB}=-\frac{\alpha}{f}\dot{\phi}\left|\bar{B}\right|^2\,,
\end{flalign}
which composes a system of equations together with Eqs. \eqref{Eq:eominflationenergydensity_BR} and \eqref{Eq:eomAenergydensity_BR} that only requires initial conditions and the input of the mean value of the magnetic field. We use the WKB approximation for the vector modes to estimate the later. Only the modes that suffer a tachyonic enhancement, $k<2\xi a H$, become classical and can contribute to source the backreaction, 
\begin{equation}
	\langle\left|\bar{B}\right|^2\rangle=\frac{1}{8\pi^3a^4}\int d^3k\, k^2\sum_{\lambda=\pm}\left|A_+\right|^2\simeq\frac{e^{2\pi\xi}}{4\pi^2a^4}\int dk\, k^3 \left(\frac{k}{2\xi a H}\right)^{1/2}e^{-4\sqrt{2\xi k/\left(aH\right)}}\simeq\frac{e^{2\pi\xi}}{\pi^2\xi}\left(\frac{H}{64\xi}\right)^4\times I[8\xi] \,,
\end{equation}
with 
\begin{equation}
	I[x]=8!\left(1-e^{-x}\right)-8!\,e^{-x}\sum_{n=1}^{8}\frac{x^n}{n!}\,.
\end{equation}
This system of equations acts as the $0^{th}$-order equations present in the gradient expansion formalism \cite{Gorbar:2021rlt}. We have taken only this first correction to a no-backreacting system as it provides a consistent evolution for any coupling $\alpha/f$ until the end of reheating. 

\begin{figure}[H]
	\centering
	\includegraphics[totalheight=6cm]{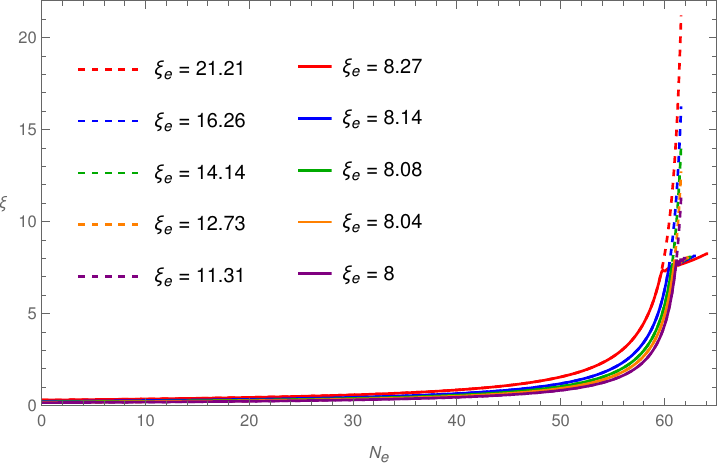} 
	\caption{Left panel: comparison of $\xi$ evolution with $0^{th}$-order (dashed) and without (solid) backreaction effects for  $\alpha m_P/f=$ 30, 23, 20 and 18 and 16. Without backreaction, this gives respectively: $\xi_{60}=$0.3, 0.23, 0.2, 0.18 and 0.16, at 60 e-folds}
	\label{fig:xievolutionbr}	
\end{figure}

Computing  $\xi$, and starting with the same initial conditions for the inflaton field than in Fig. \ref{fig:xievolutionnbr}, one now sees in Fig. \ref{fig:xievolutionbr} a flattening of the curves at the end of inflation when $\epsilon_H=1$, leading to $\xi_e<9$. In addition, one has some extra e-folds of inflation due to the slowing of the inflaton velocity. Therefore, in the case with backreaction, as we deviate from a pure single field slow-roll evolution, the definition for $\xi_{60}$ Eq. \eqref{Eq:xiCMB} no longer holds.
With the same inflaton initial conditions we now have a larger $\xi_{60}$ than in a no-backreaction regime.
Nonetheless, for an $\alpha$-attractor potential as the inflaton velocity is extremely low at the early e-folds the difference can be neglected. 

\begin{figure}[H]
  \centering
  	\includegraphics[totalheight=6.5cm]{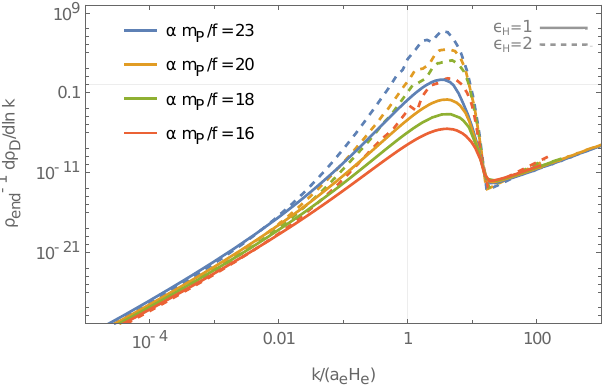} 

	\caption{Vector energy density spectrum normalized with the total energy density at the end of inflation, for different values of the inflaton-vector coupling $\alpha/f$ as indicated in the plot.  
          In solid lines one has the spectrum at $\epsilon_H=1$, whereas dashed lines are the result at $\epsilon_H=2$. }
	\label{fig:enrgydensitybr}	
\end{figure}

Although with a better picture in regards to the $\xi$ parameter, and smaller values at the end of inflation $\xi_e$,  we ought to compute again the spectral energy density to verify if the system maintains the correct energy balance.  Normalizing this quantity with the total energy density, inflaton plus vector modes, we obtain the results shown in Fig. \ref{fig:enrgydensitybr}.
Solid lines give the spectrum at the end of inflation when $\epsilon_H=1$, and for example 
for
$\alpha m_P/f=23$ the energy of the amplified vector modes will be larger than the total energy density in the universe, revealing that the backreaction effects are not still under control.

The situation is direr if one takes into consideration the reheating transition, letting the system evolve until $\epsilon_H=2$, dashed lines in Fig. \ref{fig:enrgydensitybr}. Note that having an inflaton in a $\alpha$-attractor model, reheating effectively as a quartic potential \cite{Bastero-Gil:2020uww}, we have ensured a radiation-like stage after inflation, since the scalar field oscillating about the minimum redshifts as radiation. Although modes for which $k\gtrsim aH$ at $\epsilon_H=1$  will remain inside the horizon, when analyzing the evolution, one realizes that there is still relevant amplitude growth until $\epsilon_H=2$ for these modes that started to be amplified during the slow-roll regime.  Thus, the parameter space where one can study gauge mode amplification with this system of backreaction equations is severely reduced, allowing only good estimates when $\alpha m_P/f< 16$.

Indeed, what we have shown is that including only $0^{th}$ order BR effects in the background evolution is not enough to capture the dynamics of the system near the end of inflaton and especially during the transition towards a radiation dominated universe, in particular that of the vector fluctuations. Even if we stay well within the linear regime when computing the evolution of the vector modes up to the end of inflation
($\alpha m_P/f < 23$), non-linearities are unavoidable during the transition towards $\epsilon_H=2$, and these effects will set the final shape of the transverse modes spectrum.  Accordingly, for the rest of the analysis we will take the subscript \textit{"end"} to mean at the end of reheating i.e. $\epsilon_H=2$.

We have kept an analysis with our $0^{th}$-order system of equations as it is the correction where we can completely control the numerical errors in the evolution from the beginning of inflation until the end of reheating at $\epsilon_H=2$. 
Recently there have been interesting results in managing the backreaction of the vector modes on the inflaton evolution without requiring an integration and inclusion of the effects mode by mode, see \cite{Sobol:2020lec,Gorbar:2021rlt,Durrer:2023rhc}. Nonetheless, no reheating has been included in these works. In a future work we hope to include a complete study for the gauge mode production. In regards to the present work, we will proceed with the numerical results obtained within the possible parameter space consistent with no BR up to $\epsilon_H=2$ in order to calculate the full gravitational spectrum produced from the gauge modes amplification.

\section{Power spectrum GW's}\label{secIV}
Here we study the production of gravitational waves induced by the electromagnetic modes. Focusing just on GW's we take the metric
\begin{equation}
d s^{2}=a^{2}(\tau)\left[-d \tau^{2}+\left(\delta_{i j}+h_{i j}\right) d x^{i} d x^{j}\right],
\end{equation}
where $h_i^i=h_{ij,j}=0$. The evolution of the tensor modes is given by
\begin{equation}
h_{i j}^{\prime \prime}+2 \frac{a^{\prime}}{a} h_{i j}^{\prime}-\Delta h_{i j}=\frac{2}{m_{P}^{2}} \Pi_{i j}{ }^{l m} T_{l m}^{E M} \,,
\label{Eq:h_eom}
\end{equation}
where $\Pi_{i j}^{l m}=\Pi_{l}^{i} \Pi_{m}^{\jmath}-\frac{1}{2} \Pi_{i j} \Pi^{l m}$ is the transverse traceless projector, with $\Pi_{i j}=\delta_{i j}-\partial_{i} \partial_{j} / \Delta$ and $T_{lm}^{EM}$ contains the spatial contributions of the electromagnetic stress energy tensor. Now we change into a description in momentum space, projecting the tensor modes on the positive and negative-helicity solutions
\begin{equation}
h^{i j}(\mathbf{k})=\sqrt{2} \sum_{\lambda=\pm} \epsilon_{\lambda}^{i}(\mathbf{k}) \epsilon_{\lambda}^{j}(\mathbf{k}) h_{\lambda}(\tau, \mathbf{k}) \,,
\end{equation}
and introduce the polarization tensors $\Pi_{\pm}^{i j}(\mathbf{k})=\epsilon_{\mp}^{i}(\mathbf{k}) \epsilon_{\mp}^{j}(\mathbf{k}) / \sqrt{2}$, so that $h_{\pm}(\mathbf{k})=\Pi_{\pm}^{i j}(\mathbf{k}) h_{i j}(\mathbf{k})$.
Using that ${\Pi_{\pm}^{i j} \Pi_{i j}^{l m}=\Pi_{\pm}^{l m}}$, the particular solution of Eq.  \eqref{Eq:h_eom} is given by 
\begin{flalign}
&h_{\pm}(\mathbf{k})=-\frac{2 H^{2}}{m_{P}^{2}} \int d \tau^{\prime} G_{k}\left(\tau, \tau^{\prime}\right) \tau^{\prime 2} \int \frac{d^{3} \mathbf{q}}{(2 \pi)^{3 }} \Pi_{\pm}^{l m}(\mathbf{k}) \times \nonumber \\
  &\times\left[A_{l}^{\prime}\left(\mathbf{q}, \tau^{\prime}\right)A_{m}^{\prime}\left(\mathbf{k}-\mathbf{q}, \tau^{\prime}\right)-\varepsilon_{l a b} q_{a}A_{b}\left(\mathbf{q}, \tau^{\prime}\right) \varepsilon_{m c d}\left(k-q\right)_{c} A_{d}\left(\mathbf{k}-\mathbf{q}, \tau^{\prime}\right)\right] \,,
\label{Eq:1st_h_modes}
\end{flalign}
where $G_{k}\left(\tau, \tau^{\prime}\right)$  is the retarded Green function for the operator $d^{2} / d \tau^{2}-(2 / \tau) d / d \tau+k^{2}$, 
\begin{equation}
G_{k}\left(\tau, \tau^{\prime}\right)=\frac{1}{k^{3} \tau^{\prime 2}}\left[\left(1+k^{2} \tau \tau^{\prime}\right) \sin k\left(\tau-\tau^{\prime}\right)+k\left(\tau^{\prime}-\tau\right) \cos k\left(\tau-\tau^{\prime}\right)\right] |
\end{equation}
for $\tau>\tau^{\prime}$, while $G_{k}\left(\tau<\tau^{\prime}\right)=0$.

We will be interested in promoting the gauge modes to operators, see Eq. \eqref{Eq:Gauge_expansion}, to then proceed with the tensor modes. As discussed in section \ref{secII} we will describe the vector mode amplitudes with a step function. In order to relate each stage to the amplitudes at the maximum, at the end of reheating,  recovering Eq. \eqref{Eq:stepfunction}, we write our step function for $A_+$ and $A'_+$ with the transfer functions $T$ and $\bar{T}$ as 
\begin{equation}
A_+(k,\tau)=A_{end}(k)T^\textbf{k}(\tau,\tau_{end})\simeq \begin{cases}
{A_{BD}(k)} & \tau<\tau_{tac}\\
{A_{WKB}(k,\tau)} & \tau_{tac}<\tau<\tau_h\\
A_{end}(k) & \tau> \tau_h\ .
\end{cases}\label{Eq:stepfunctionwithT}
\end{equation}

As previously discussed, the Bunch-Davis vacuum contributions give a UV divergence on the amplitudes that must be regularized. When integrating over all momentum $k$ and for the entire evolution of $\tau$ one automatically includes modes that will always remain sub horizon and that will not be amplified, i.e. where the condition $\tau> \tau_{tac}$ is never realized before the end of inflation. To circumvent this issue, as the tachyonic amplification effects dominate the system, when integrating the GW amplitude we remove the Bunch-Davis vacuum contribution in the gauge amplitudes. We will keep the WKB amplitude until horizon crossing. Specifically we will we use the smooth functions
\begin{flalign}
\left|A_{end}(k)\right|^2T^\textbf{k}(\tau,\tau_{end})^2 =&\frac{\left|A_{WKB}(k,\tau)\right|}{2}^2 \left[1- \tanh \left(\delta \left(\frac{\tau}{\tau_{h}}-1\right)\right)\right] \nonumber
\\+&\frac{\left|A_{end}(k)\right|}{2}^2 \left[1+\tanh \left(\delta \left(\frac{\tau}{\tau_h}-1\right)\right)\right] \,,
\label{Eq:TransferfunctionT} 
\end{flalign}
with $\delta=10$, $\tau_{tac}=-\xi/k$ and $\tau_h=-1/(10\, k)$.
An analogous description is used for $A'_+(k,\tau)$ with $\bar{T}$ as the respective transfer function. 

Furthermore, to describe the spectrum for $|A'_{end}\left( {k}\right)|^2$, $k^2|A_{end}\left( {k}\right)|^2$ and $k|AA_{end}^{\prime}\left( {k}\right)|$ at $\epsilon_H=2$, we use the functions in Eqs. \eqref{Eq:A'2spectrum}, \eqref{Eq:A2spectrum} and \eqref{Eq:AA'spectrum} which have a very good agreement with the  numerically computed spectrums at the end of reheating (see Fig. \ref{fig:amplitudes} in Appendix \ref{appendixA}):
\begin{flalign}
&|A'_{end}\left( {k}\right)|^2\simeq (a_e H_e)\exp\left[\left(x^{A'}_{0}+ x^{A'}_{1}\,\tilde{k}+ x^{A'}_{2}\,\tilde{k}^2+x^{A'}_{3}\,\tilde{k}^3\right)\left(1-e^{a^{A'}_{0} (\tilde{k}-a^{A'}_{1})}\right)\right] \,, \label{Eq:A'2spectrum}
\\
&k^2 |A_{end}\left( {k}\right)|^2\ \simeq(a_e H_e)\exp\left[\left(x^{A}_{0}+ x^{A}_{1}\,\tilde{k}+ x^{A}_{2}\,\tilde{k}^2+x^{A}_{3}\,\tilde{k}^3\right)\left(1-e^{a^{A}_{0} (\tilde{k}-a^{A}_{1})}\right)\right]  \,, \label{Eq:A2spectrum}
\\
&k \left|AA'_{end}({k}) \right|\simeq(a_e H_e)\exp\left[\left(x^{AA'}_{0}+ x^{AA'}_{1}\,\tilde{k}+ x^{AA'}_{2}\,\tilde{k}^2+x^{AA'}_{3}\,\tilde{k}^3\right)\left(1-e^{a^{AA'}_{0} (\tilde{k}-a^{AA'}_{1})}\right)\right]  \,, \label{Eq:AA'spectrum}
\end{flalign}  
with $\tilde{k}=k/(a_eH_e)$ and where the $x$'s and $a$'s are functions of $\alpha\,m_P/f$ listed in Appendix \ref{appendixA}. Furthermore, as visible in  Fig. \ref{fig:amplitudes}, the use of the functions in Eqs. \eqref{Eq:A'2spectrum}, \eqref{Eq:A2spectrum} and \eqref{Eq:AA'spectrum} provides us a regulator that controls, in a semi-analytical way, the divergences that come from the higher momentum vacuum contributions. Finally, setting $A_-=0$ we calculate the two point function for the gravitational waves as described in detail in Appendix \ref{appendixB}.

\begin{figure}[H]
	\centering
	\includegraphics[totalheight=6.5cm]{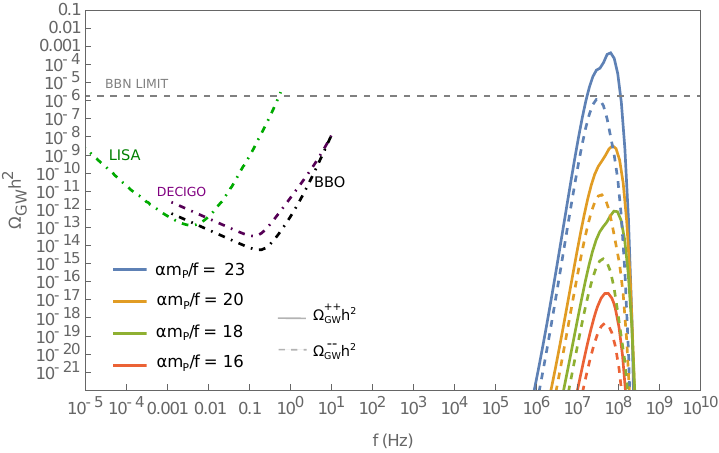}
	\caption{Spectral  density of GW's in frequency today sourced by the tachyonic amplification of vector modes. Sensitivity curves for LISA, DECIGO and BBO are included as well as the BBN energy density limit.}
	\label{fig:spectrumGWS}	
\end{figure}

\begin{figure}[H]
	\centering
	\includegraphics[totalheight=5.6cm]{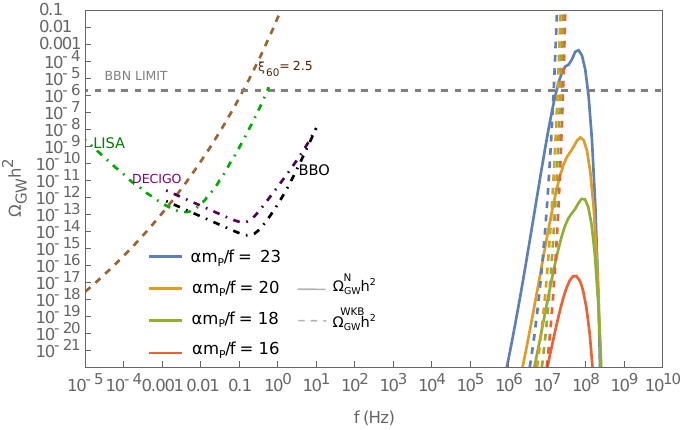} \includegraphics[totalheight=5.6cm]{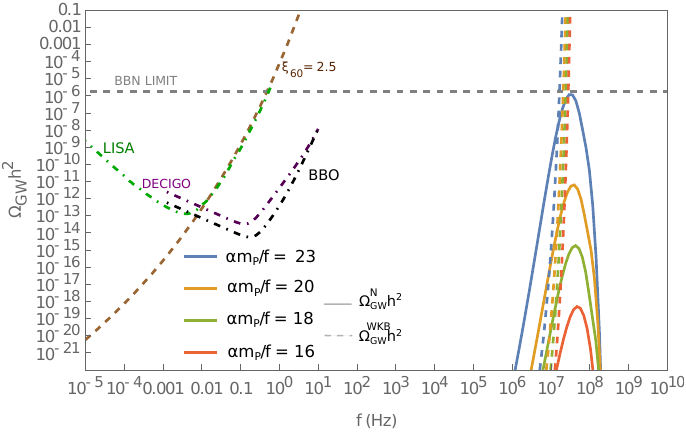}
	\caption{Comparison of the spectral energy density for the GW's produced from the tachyonic amplification estimated by the semi analytical method described with Eqs. \eqref{Eq:final_2point_h_modes} and \eqref{Eq:GWspectrum}, and the analytical description with the WKB solution for the vector modes Eqs. \eqref{Eq:GW--WBK} and \eqref{Eq:GW++WBK}. In the left panel we show the $++$ two point function and in the right panel the $--$ correlation.}
	\label{fig:spectrumGWSvswkb}	
\end{figure}

Taking $x=-k\tau$ and $\tilde{q}=q/(a_eH_e)$ we get
\begin{flalign}
\left\langle h_{s}h_{s'}\right\rangle 	=&\frac{2}{(2\pi)^3}\frac{H^{4}}{{m_P}^{4}}\frac{\left(a_{e}H_{e}\right){}^{5}}{k^{8}}\text{\ensuremath{\int_{0}^{\infty}\int_{0}^{\infty}}d\ensuremath{x_{1}}d\ensuremath{x_{2}\left(\sin x_{1}-x_{1}\cos x_{1}\right)\left(\sin x_{2}-x_{2}\cos x_{2}\right)}}\nonumber
\\&\int_{0}^{\infty}d\tilde{q}\int_{0}^{2\pi}\frac{2\pi}{16}\tilde{q}^{2}d\theta(1+s\cos\theta)\left(1+s'\cos\theta\right)\left(1+s\frac{1-\frac{\tilde{q}\cos\theta}{\tilde{k}}}
{\sqrt{1+\frac{\tilde{q}^2}{\tilde{k}^2}-2\frac{\tilde{q}}{\tilde{k}}\cos\theta}}\right)\left(1+s'\frac{1-\frac{\tilde{q}\cos\theta}{\tilde{k}}}{\sqrt{1+\frac{\tilde{q}^2}{\tilde{k}^2}-2\frac{\tilde{q}}{\tilde{k}}\cos\theta}}\right)\nonumber
\\&(\tilde{q}^2(\tilde{k}-\tilde{q})^{2}\left|A_{end}\left(\tilde{q}\right)\right|^{2}\left|A_{end}\left(\tilde{k}-\tilde{q}\right)\right|^{2}T_{\tilde{q}}^{x_{1}}T_{\tilde{q}}^{x_{2}}T_{\tilde{k}-\tilde{q}}^{x_{1}}T_{\tilde{k}-\tilde{q}}^{x_{2}}\nonumber
\\&+\left|A_{end}'\left(\tilde{q}\right)\right|^{2}\left|A_{end}'\left(\tilde{k}-\tilde{q}\right)\right|^{2}\bar{T}_{\tilde{q}}^{x_{1}}\bar{T}_{\tilde{q}}^{x_{2}}\bar{T}_{\tilde{k}-\tilde{q}}^{x_{1}}\bar{T}_{\tilde{k}-\tilde{q}}^{x_{2}}\nonumber
\\&+\tilde{q}(\tilde{k}-\tilde{q})\left|AA_{end}'\left(\tilde{q}\right)\right|\left|AA_{end}'\left(\tilde{k}-\tilde{q}\right)\right|(T_{\tilde{q}}^{x_{1}}\bar{T}_{\tilde{q}}^{x_{2}}T_{\tilde{k}-\tilde{q}}^{x_{1}}\bar{T}_{\tilde{k}-\tilde{q}}^{x_{2}}+T_{\tilde{q}}^{x_{2}}\bar{T}_{\tilde{q}}^{x_{1}}T_{\tilde{k}-\tilde{q}}^{x_{2}}\bar{T}_{\tilde{k}-\tilde{q}}^{x_{1}}))\,.
	\label{Eq:final_2point_h_modes}
\end{flalign}
From the correlation functions we obtain the power spectrum that is then related to the fraction of energy density of gravitational waves today by \cite{Guzzetti:2016mkm,Caprini:2018mtu}
\begin{equation}
	\Omega_{ss'}h^2=\frac{\Omega_{R\,0}h^2}{24}\frac{k^3}{2\pi^2}\left\langle h_{s}h_{s'}\right\rangle \,, \label{Eq:GWspectrum}
\end{equation}
where $\Omega_{R\,0}h^2\equiv\rho_{R\,0}h^2/3H_0^2m_P^2\simeq4.18\times10^{-5}$. Finally the dependence in $k$ is related to the frequency today by
\begin{equation}
	f= \frac{k}{2\pi a_0}= \frac{\tilde{k}}{2\pi}H_e\frac{a_e}{a_0} \,,
\end{equation}
with $\tilde{k}=k/(a_eH_e)$. In Fig. \ref{fig:spectrumGWS} we represent the spectral energy density of the induced gravitational waves from Eq. \eqref{Eq:GWspectrum} for both $s=s'=+$ (solid lines) and $s=s'=-$ (dashed lines), and several values of $\xi_{60}$. 
The BBN limit, $\Omega_{GW} h^2<1.8\times10^{-6}$, that sets an upper bound on the radiation excess at BBN, and the sensitivity curves for planned GW detectors are also included \cite{Yagi:2011wg,Bartolo:2016ami}. 
We find a parity asymmetric spectrum with a difference at the peaks of 2/3 orders of magnitude between the $++$ and $--$ correlations, in line with the results obtained at CMB scales in \cite{Sorbo:2011rz,Garcia-Bellido:2016dkw,Bartolo:2016ami}. 
The peaks for the spectral distribution come around {$10^7,10^8\ \mathrm{Hz}$}, typical in end of inflation and (p)reheating stages, exactly where we had the maximum spectral amplification for the vector modes. 
At larger frequencies the spectrum falls exponentially, again following the effects of the electromagnetic sources with a correct vacuum subtraction \cite{Parker:2009uva}.  
In the range of frequencies where one could have a detection of the represented signals there are no current or planed GW detectors, although interesting proposals and motivations  are discussed in \cite{GWMHz,Aggarwal:2020umq,Herman:2020wao,Giovannini:2023itq,Domcke:2023qle}.
At present and planed interferometer scales the energy densities are negligible  for the interaction parameters $\alpha m_P/f$ we were able to consider.  
For $\alpha m_P/f=23$, a parameter already beyond the linear backreaction treatment that we have employed as discussed in section \ref{secIII}, we obtain a energy density that surpasses the BBN bound. 
This does not exclude models with this or larger couplings, only reveals the need for an appropriate description of the vector mode amplification, fully including non-linearities in the evolution. Moreover, from the figure one can also see that production in the linear regime $\alpha m_P/f<16$, for an $\alpha$-attractor potential, will lead to an extremely small signal in the GW spectrum. The spectrum going beyond the linear regime remains to be calculated. 
 We now compare, in Fig. \ref{fig:spectrumGWSvswkb}, our results with the expressions obtained in \cite{Bartolo:2016ami}, following the work in \cite{Sorbo:2011rz}, where the analytic expression in Eq. \eqref{Eq:WKB_vector} was employed on the two point correlation function.
\begin{flalign}
\Omega^{WKB}_{++}h^2\simeq1.5\times 10^{-13}\frac{H^4}{m_P^4}\frac{e^{4\pi\xi}}{\xi^6}\,, \label{Eq:GW++WBK}\\
\Omega^{WKB}_{--}h^2\simeq3.1\times 10^{-16}\frac{H^4}{m_P^4}\frac{e^{4\pi\xi}}{\xi^6}\,. \label{Eq:GW--WBK}
\end{flalign}
The shortcomings of the analytical estimations with the WKB expansion are apparent in Fig.  \ref{fig:spectrumGWSvswkb}. With the expressions in Eqs. \eqref{Eq:GW--WBK} and \eqref{Eq:GW++WBK} one may get appropriate descriptions at CMB scales, allowing for instance the study of parity violating signals in the B-modes \cite{Sorbo:2011rz}. However, at frequencies typical of end of inflation (MHz) the expressions become less reliable as we are no longer in the constant $\xi$ regime and the deviations from our curves are substantial. Furthermore, for an interaction parameter $\xi_{60}\simeq 2.5$
$(\alpha m_{P}/f\simeq249)$,
as was considered in \cite{Bartolo:2016ami}, the linear description no longer holds and the estimation for a detection with LISA is a stretch for the model capabilities.  Nonetheless, as the non-linear dynamics present with strong backreaction remains to be integrated in the GW spectrum calculation, the possibility may be dim but it is not excluded.

\begin{figure}[H]
	\centering
	\includegraphics[totalheight=6.5cm]{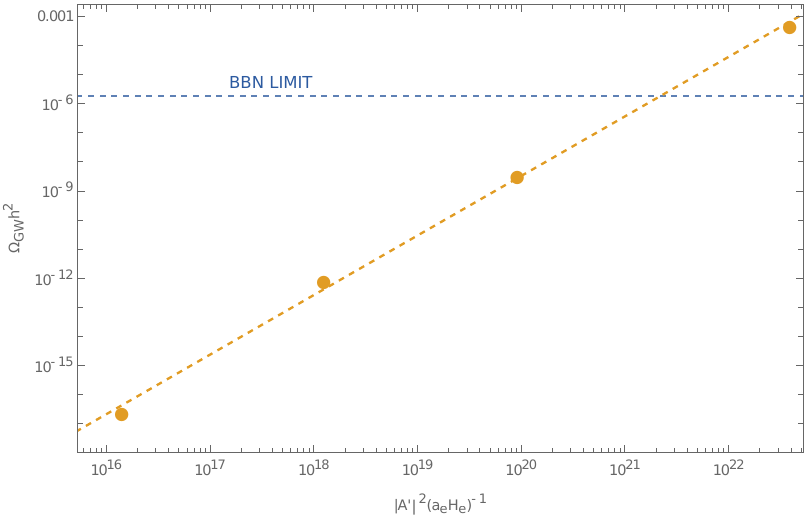}
	\caption{Linear interpolation to obtain the maximum amplitude on the electric field contribution $|A'(k)|^2$. The points to obtain the linear fit were obtained using (from left to right) $\alpha m_P/f=$16, 18, 20, 23.}
	\label{fig:fitBBNbound}	
\end{figure}

Finally, we can combine the numerical procedure to obtain the energy density of the  gravitational waves today, where we use the amplitude spectrum of the gauge modes contributions through Eqs. \eqref{Eq:A2spectrum}, \eqref{Eq:A'2spectrum} and \eqref{Eq:AA'spectrum}, and the upper bound on the radiation excess at BBN to set an estimate of a ceiling on the amplitude of such electromagnetic sources. 
As each individual contribution, from $|A'(k)|^2$, $k^2|A(k)|^2$ and $k|A'(k)A(k)|$, are of similar order, through a simple linear fit we find the would be maximum amplitude of the electric contribution $|A'(k)|^2$, our representative, to reach the BBN bound, see Fig.  \ref{fig:fitBBNbound}. We obtain a maximum amplitude at $|A'(k)|^2\sim10^{21}\, a_e H_e$.
Note that this analysis is independent of a correct or incorrect parametrization of the backreaction effects on the inflaton evolution. The assumption is that the gauge modes spectrum will keep a similar shape, where the main contribution for the integration in Eq. \eqref{Eq:final_2point_h_modes} comes through a dominant peak for the modes amplified close to the end of inflation as in the amplitudes represented in Fig. \ref{fig:amplitudes}. One might expect a broadening of the vector mode spectrum with a correct account of the backreaction effects leading to larger contribution of the smaller $k$ modes, thus this result will provide an upper bound on the maximum amplitude of the electric contribution. As we will present in Appendix \ref{appendixC}, this relation may also depend on the inflationary potential as the shape of the vector modes is influenced by the inflaton velocity.

\section{Conclusion}\label{secV}

In this work we have calculated the gravitational wave spectrum produced from the ``axion-like'' interaction between the inflaton and an $U(1)$ gauge field, in the linear regime during inflation and reheating. Main results are presented in Fig.  \ref{fig:spectrumGWS}. 

During the slow-roll evolution, the inflaton motion sources a tachyonic amplification of the gauge field amplitudes. From the asymmetry present in the $\phi F\tilde{F}$ interactions, only one of the transverse polarizations is amplified resulting in a parity asymmetric source for the gravitational waves. Naturally this propagates into the GW spectrum as seen in Fig. \ref{fig:spectrumGWS} and predicted when contrasting Eqs. \eqref{Eq:GW++WBK} and \eqref{Eq:GW--WBK}. The peaks in the GW spectrum appear around $10^8$ Hz, as expected from the maximal amplification of the vector modes at the end of inflation and reheating. We have also shown an example with $\alpha m_{P}/f\simeq23$ ($\xi_{60}\simeq 0.23$) , where computing a spectrum with a inadequate description could point to an erroneous exclusion of a parameter space due to a crossing into the BBN limit on the GW energy density. The analytical predictions obtained with the WKB solution Eqs. \eqref{Eq:GW++WBK} and \eqref{Eq:GW--WBK} vary significantly from our curves at large frequencies, as seen in Fig. \ref{fig:spectrumGWSvswkb}.
The prediction of a detection with LISA obtained for $\xi_{60}\simeq 2.5$ $(\alpha m_P/f\simeq 249)$ goes beyond the validity of the WKB analytical description. Nevertheless, a signal within the detector sensitivity may come as a combination of the undescribed non-linear dynamics in the strong backreaction regime and a low scale inflation model compatible with the observations\footnote{For instance, with the potential in \cite{Takahashi:2018tdu} one may lower inflation scale enough to possibly generate a peak around Hz scale.}.
At the large frequencies ranges (MHz) predicted in our calculations there are no planned or expected detectors. Nonetheless there is an a raising interest in studying such scales, see \cite{GWMHz,Aggarwal:2020umq,Herman:2020wao,Giovannini:2023itq,Domcke:2023qle}.
We thus hope that this work motivates proposals and further development of ideas to detect signals within this regime where there are no known astrophysical sources.

To derive the GW spectrum we have studied the vector production until the end of reheating when the tachyonic amplification comes to a halt. The gauge field amplitudes are described with a smoothed step function, initially with the analytical WKB solution, and after horizon crossing with the amplitudes value at $\epsilon_H=2$, obtained through the numerical integration of the equations of motion. Therein we combine the good estimation in the WKB solution for the start of the tachyonic enhancement with an almost constant amplitude from the horizon crossing of the modes until the end of the inflationary dynamics. Here we have considered a simple system to attempt to mimic the backreaction effects on the inflationary motion based on energy conservation on the vector modes. We were then able to reproduce the vector production with the correct description for $\alpha m_P/f< 16$ . However in this linear regime, signals in the GW spectrum are extremely small. Furthermore, we have obtained an estimation of the upper bound on the amplitudes of the electromagnetic sources, $k^2|A(k)|^2$, $|A'(k)|^2$, $k |A'(k)A(k)|$. In order to avoid radiation excess at BBN one has to verify $|A'(k)|^2\lesssim10^{21}\, a_e H_e$.

In order to extend the parameter space to the constrains given by the upper bound on non-Gaussianities, ${\xi_{60}\simeq2.5}$ $(\alpha m_P/f\simeq 249)$, the non-linear dynamics of the backreaction effects have to be integrated in the system. With the gradient expansion formalism, through a system of 3-n differential equations for bilinear functions of the electromagnetic fields in coordinate space, the authors of \cite{Gorbar:2021rlt}  were able to manage those effects until the end of inflation, $\epsilon_H=1$ for a quadratic inflationary potential. The oscillating effects in the $\xi$ evolution close to the end of inflation, also confirmed in the works \cite{Cheng:2015oqa,DallAgata:2019yrr,Domcke:2020zez,Caravano:2022epk,Peloso:2022ovc,Figueroa:2023oxc}, seem to induce a double peak in the gauge particle amplitudes spectrum that could result in interesting effects on the GW spectrum. 

As future a direction of this work  we will be looking for a correct estimation of vector production in the strong backreaction case both during inflation and until the end of reheating, at $\epsilon_H=2$. To then estimate the gravitational wave spectrum in the entirety of the allowed parameter range. It would also be interesting to study if the GW spectrum exhibits non-Gaussian statistics inducing a more distinct signal on a possible detection.

\appendix

\section{Spectrum for $E^2$, $B^2$ and $E\cdot B$}\label{appendixA}

\begin{figure}[H]
	\centering
	\includegraphics[totalheight=7cm]{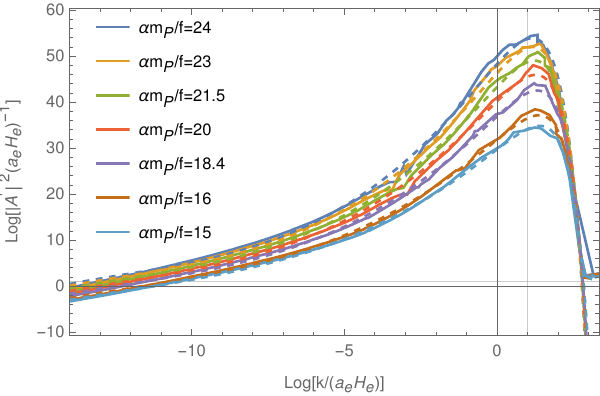}
	\includegraphics[totalheight=7cm]{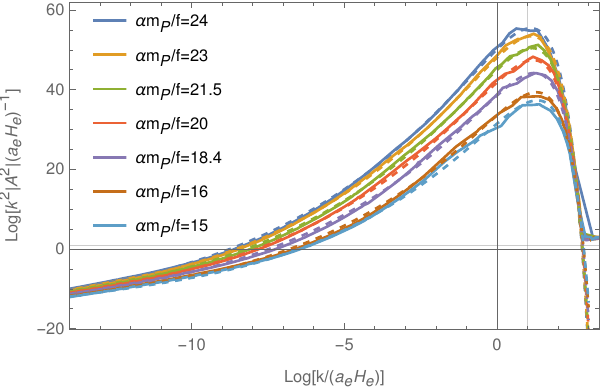}
	\includegraphics[totalheight=7cm]{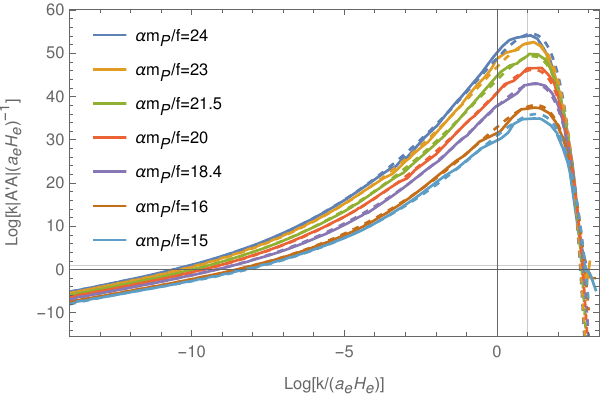}
	\caption{Spectrum for $E^2$ ($|A'(k)|^2$), $B^2$ ($k^2|A(k)|^2$) and $E\cdot B$ ($k|A'(k)A(k)|$) contributions at $\epsilon_H=2$  compared to functions in Eqs. \eqref{Eq:A'2spectrum}, \eqref{Eq:A2spectrum} and \eqref{Eq:AA'spectrum}, respectively. We have considered different values for the inflaton-vector coupling $\alpha m_P/f$.}
	\label{fig:amplitudes}	
\end{figure}

We present here how we modeled the parameters in the functions in Eqs. \eqref{Eq:A'2spectrum}, \eqref{Eq:A2spectrum} and \eqref{Eq:AA'spectrum} in terms of ${z=\alpha m_P/f}$.  
In Fig. \eqref{fig:amplitudes} we compared our semi-analytical parametrization with the spectrum obtained numerically for the amplitudes of $a^4E^2$, $a^4B^2$ and $a ^4E\cdot B$.
\begin{flalign}
&x^{A'}_0(z)=-2.74794 + 2.1837\, z\\
&x^{A'}_1(z)=-0.772418 + 0.422345\,  z \\
&x^{A'}_2(z)=-0.128614 + 0.033448 \, z \\
&x^{A'}_3(z)=-0.00304119 + 0.000889556\, z\\
&a^{A'}_0(z)=2.1696 - 0.0366212\, z \\
&a^{A'}_1(z)=2.94837 - 0.00690561 \, z  
\end{flalign} 
\begin{flalign}
&x_0^{A}(z)=-0.0253472 + 2.15692\, z\\
&x_1^{A}(z)=4.3119 + 0.226113 \, z \\
&x_2^{A}(z)=0.447448 + 0.00360117 \, z \\
&x_3^{A}(z)=0.0153263 - 0.000179546\, z\\
&a_0^{A}(z)=1.36358 - 0.00473288 \, z \\
&a_1^{A}(z)=2.93832 - 0.00626591 \, z 
\end{flalign} 
\begin{flalign}
&x_0^{AA'}(z)=1.78408 + 2.01315 \, z\\
&x_1^{AA'}(z)=3.46606 + 0.247637\, z \\
&x_2^{AA'}(z)=0.413299 + 0.00765184 \, z \\
&x_3^{AA'}(z)=0.0171652 - 0.000100337\, z\\
&a_0^{AA'}(z)=0.91585 + 0.0183234 \, z \\
&a_1^{AA'}(z)=3.18199 - 0.0187605 \, z 
\end{flalign}

\section{Details on $\left\langle h_{s}h_{s'}\right\rangle $}\label{appendixB}
We want to calculate the two point functions trying to keep the result in terms of the amplitude solutions for the gauge field modes at the end of inflation. Let us recover the vector mode expansion
\begin{equation}
\hat{A}_{i}(\tau, \mathbf{x})=\int \frac{d^{3} \mathbf{k}}{(2 \pi)^{3 }} e^{i \mathbf{k} \cdot \mathbf{x}} \hat{A}_{i}(\tau, \mathbf{k})=\sum_{\lambda=\pm} \int \frac{d^{3} \mathbf{k}}{(2 \pi)^{3  }}\left[\epsilon_{\lambda}^{i}(\mathbf{k}) A_{\lambda}(\tau, \mathbf{k}) \hat{a}_{\lambda}^{\mathbf{k}} e^{i \mathbf{k} \cdot \mathbf{x}}+\mathrm{h.c.}\right]
\end{equation}
where $\hat{a}_{\lambda}^{\mathbf{k}}$, $\hat{a}_{\lambda}^{\mathbf{k}\dagger}$ are the annihilation and creation operators.
Now, we promote the tensor modes to operators
\begin{flalign}
\hat{h}_{\pm}(\mathbf{k})&=-\frac{2\, H^{2}}{\sqrt{2}\, m_{P}^{2}} \int d \tau^{\prime} G_{k}\left(\tau, \tau^{\prime}\right) \tau^{\prime 2} \int \frac{d^{3} \mathbf{q}}{(2 \pi)^{3 }} \times \nonumber \\
&\times\left[\hat{\mathcal{A}}_{\pm, \lambda}^{\prime}\left(\mathbf{k},\mathbf{q}, \tau^{\prime}\right)\hat{\mathcal{A}}_{\pm, \lambda'}^{\prime}\left(\mathbf{k},\mathbf{k}-\mathbf{q}, \tau^{\prime}\right)-\varepsilon_{l a b} q_{a}\hat{\mathcal{A}}_{\pm, \lambda}\left(\mathbf{k},\mathbf{q}, \tau^{\prime}\right) \varepsilon_{m c d}\left(k-q\right)_{c} \hat{\mathcal{A}}_{\pm, \lambda'}\left(\mathbf{k},\mathbf{k}-\mathbf{q}, \tau^{\prime}\right)\right]\\
&=-\frac{2\, H^{2}}{\sqrt{2}\,m_{P}^{2}} \int d \tau^{\prime} G_{k}\left(\tau, \tau^{\prime}\right) \tau^{\prime 2} \int \frac{d^{3} \mathbf{q}}{(2 \pi)^{3 }}  \times \nonumber\\
&\sum_{\lambda, \lambda'=\pm}\left[\left(\epsilon^l_\mp(\textbf{k})\epsilon^l_\lambda(\textbf{q})A_{\lambda}^{\prime}\left(\mathbf{q}, \tau^{\prime}\right)\hat{a}_{\lambda}^{\mathbf{k}}+h.c.\right)
\left(\epsilon^m_\mp(\textbf{k})\epsilon^m_{\lambda'}(\mathbf{k}-\mathbf{q}) A_{\lambda'}^{\prime}\left(\mathbf{k}-\mathbf{q}, \tau^{\prime}\right)\hat{a}_{\lambda}^{\mathbf{k}}+h.c.\right)\right.\nonumber\\
& \left. -(-\lambda)(-\lambda')i^2q\left|k-q\right|\left(\epsilon^l_\mp(\textbf{k})\epsilon^l_\lambda(\textbf{q})A_{\lambda}\left(\mathbf{q}, \tau^{\prime}\right)\hat{a}_{\lambda}^{\mathbf{k}}+h.c.\right)
\left(\epsilon^m_{\mp}(\mathbf{k})\epsilon^m_{\lambda'}(\mathbf{k}-\mathbf{q})A_{\lambda'}\left(\mathbf{k}-\mathbf{q}, \tau^{\prime}\right)\hat{a}_{\lambda}^{\mathbf{k}}+h.c.\right)\right]
\label{Eq:2nd_h_modesB}
\end{flalign}
where $\hat{\mathcal{A}}_{s,\lambda}\left(\mathbf{k},\mathbf{q}, \tau^{\prime}\right)=\epsilon^l_{-s}(\textbf{k})\epsilon^l_\lambda(\textbf{q})A_{\lambda}\left(\mathbf{q}, \tau^{\prime}\right)\hat{a}_{\lambda}^{\mathbf{k}}$, we have used $\varepsilon_{a b c} k_{b} \epsilon_{\lambda}^{c}=-\lambda i k \epsilon_{\lambda}^{a}$ and decomposed $\Pi_{\pm}^{l m}(\mathbf{k})$.
Now, to calculate the two point function we may use Wick's theorem to simplify our expression, the only non zero terms will be given by $\langle \hat{c}_{\lambda}^{\mathbf{k}} \hat{c}_{\lambda'}^{\mathbf{k'}\dagger}\rangle=(2\pi)^3\delta\left(\textbf{k}-\textbf{k}'\right)\delta_{\lambda \lambda'}$ contributions
\begin{flalign}
\left\langle{ h_{s}}(\mathbf{k}) h_{s'}\left(\mathbf{k}^{\prime}\right)\right\rangle=&  2\frac{H^{4} }{ m_{P}^{4}} \int d \tau^{\prime} d \tau^{\prime \prime}  \tau^{\prime 2}  \tau^{\prime\prime 2} G_{k}\left(\tau, \tau^{\prime}\right) G_{k}\left(\tau, \tau^{\prime \prime}\right) \int \frac{d^{3} \mathbf{q}}{(2 \pi)^{3 }}\int \frac{d^{3} \mathbf{q'}}{(2 \pi)^{3 }}\sum_{\lambda, \lambda',\varLambda,\varLambda'}^\pm\nonumber \\
& \langle\left[\epsilon^l_{-s}(\textbf{k})\epsilon^l_\lambda(\textbf{q})\epsilon^m_{-s'}(\textbf{k})\epsilon^m_{\lambda'}(\mathbf{k}-\mathbf{q})
A_{\lambda}^{\prime}\left(\mathbf{q}, \tau^{\prime}\right) A_{\lambda'}^{\prime}\left(\mathbf{k}-\mathbf{q}, \tau^{\prime}\right)\hat{a}_{\lambda}^{\mathbf{q}}
\hat{a}_{\lambda'}^{\mathbf{k-q}}\right.\nonumber\\
&\left.+\lambda\lambda'q(k-q)\epsilon^l_{-s}(\textbf{k})\epsilon^l_\lambda(\textbf{q})\epsilon^m_{-s'}(\textbf{k})\epsilon^m_{\lambda'}(\mathbf{k}-\mathbf{q})
A_{\lambda}\left(\mathbf{q}, \tau^{\prime}\right)A_{\lambda'}\left(\mathbf{k}-\mathbf{q}, \tau^{\prime}\right)\hat{a}_{\lambda}^{\mathbf{q}}\hat{a}_{\lambda'}^{\mathbf{k-q}} \right]\nonumber\\
& \left[\epsilon^{*a}_{-s}(\textbf{k}')\epsilon^{*a}_\varLambda(\textbf{q}')\epsilon^{*b}_{-s'}(\textbf{k}')\epsilon^{*b}_{\varLambda'}(\mathbf{k'}-\mathbf{q'})
A_{\varLambda}^{*\prime}\left(\mathbf{q'}, \tau^{\prime\prime}\right) A_{\varLambda'}^{*\prime}\left(\mathbf{k'}-\mathbf{q'}, 	\tau^{\prime\prime}\right)
\hat{a}_{\varLambda}^{\dagger\mathbf{q'}}\hat{a}_{\varLambda'}^{\dagger\mathbf{k'-q'}}\right.\nonumber\\
&
+\left.\varLambda\varLambda'q'(k'-q')\epsilon^{*a}_{-s}(\textbf{k}')\epsilon^{*a}_\varLambda(\textbf{q}')\epsilon^{*b}_{-s'}(\textbf{k}')\epsilon^{*b}_{\varLambda'}(\mathbf{k'}-\mathbf{q'})
A^*_{\varLambda}\left(\mathbf{q'}, \tau^{\prime\prime}\right)A^*_{\varLambda'}\left(\mathbf{k'}-\mathbf{q'}, 	\tau^{\prime\prime}\right)\hat{a}_{\varLambda}^{\dagger\mathbf{q'}}\hat{a}_{\varLambda'}^{\dagger\mathbf{k'-q'}} \right] \rangle
\label{Eq:1st_2point_h_modesB}
\end{flalign} 
with the relations on $\langle \hat{a}_{\lambda}^{\mathbf{q}}  \hat{a}_{\lambda'}^{\mathbf{k}-\mathbf{q}} \hat{a}_{\varLambda}^{\dagger\mathbf{q'}}\hat{a}_{\varLambda'}^{\dagger\mathbf{k'}-\mathbf{q'}}\rangle$ the two point function can be simplified into
\begin{flalign}
\left\langle{ h_{s}}(\mathbf{k}) h_{s'}\left(\mathbf{k}^{\prime}\right)\right\rangle= & \frac{2}{(2\pi)^3}\frac{H^{4} }{ m_{P}^{4}} \delta\left(\mathbf{k}+\mathbf{k}^{\prime}\right)\int d \tau^{\prime} d \tau^{\prime \prime}  \tau^{\prime 2}  \tau^{\prime\prime 2} G_{k}\left(\tau, \tau^{\prime}\right) G_{k}\left(\tau, \tau^{\prime \prime}\right)\nonumber\\
& \int d^{3} \mathbf{q}\sum_{\lambda, \lambda'}^\pm \epsilon_{-s}^{l}(\mathbf{k}) \epsilon^l_\lambda(\textbf{q})\epsilon_{-s}^{m}(\mathbf{k})\epsilon^m_{\lambda'}(\mathbf{k}-\mathbf{q})\epsilon_{-s'}^{*a}(\mathbf{k}) \epsilon^{*a}_\lambda(\textbf{q})\epsilon_{-s'}^{*b}(\mathbf{k}) \epsilon^{*b}_{\lambda'}(\mathbf{k}-\mathbf{q}) \nonumber \\
& \times\left[
A_{\lambda}^{\prime}\left(\mathbf{q}, \tau^{\prime}\right) A_{\lambda'}^{\prime}\left(\mathbf{k}-\mathbf{q}, \tau^{\prime}\right)
+\lambda\lambda'q(k-q)A_{\lambda}\left(\mathbf{q}, \tau^{\prime}\right)A_{\lambda'}\left(\mathbf{k}-\mathbf{q}, \tau^{\prime}\right)\right]\nonumber\\
& \times\left[
A_{\lambda}^{*\prime}\left(\mathbf{q}, \tau^{\prime\prime}\right) A_{\lambda}^{*\prime}\left(\mathbf{k}-\mathbf{q},\tau^{\prime\prime}\right)+\lambda\lambda'q(k-q)
A^*_{\lambda}\left(\mathbf{q}, \tau^{\prime\prime}\right)A^*_{\lambda'}\left(\mathbf{k}-\mathbf{q}, \tau^{\prime\prime}\right) \right]. 
\label{Eq:2nd_2point_h_modesB}
\end{flalign}
As studied for the cosmological evolution during inflation, the gauge modes with $+$ helicity will be severely amplified and the negative solution will be suppressed. Taking $A_-\simeq 0$ we find 
\begin{flalign}
&\left\langle{ h_{s}}(\mathbf{k}) h_{s'}\left(\mathbf{k}^{\prime}\right)\right\rangle=  \frac{2}{(2\pi)^3}\frac{H^{4} }{ m_{P}^{4}} \delta\left(\mathbf{k}+\mathbf{k}^{\prime}\right)\int d \tau^{\prime} d \tau^{\prime \prime}  \tau^{\prime 2}  \tau^{\prime\prime 2} G_{k}\left(\tau, \tau^{\prime}\right) G_{k}\left(\tau, \tau^{\prime \prime}\right)\nonumber\\
& \int d^{3} \mathbf{q}\ \epsilon_{-s}^{l}(\mathbf{k}) \epsilon^l_+(\textbf{q})\epsilon_{-s'}^{*a}(\mathbf{k}) \epsilon^{*a}_+(\textbf{q}) \epsilon_{-s}^{m}(\mathbf{k})\epsilon^m_{+}(\mathbf{k}-\mathbf{q})\epsilon_{-s'}^{*b}(\mathbf{k})\epsilon^{*b}_{+}(\mathbf{k}-\mathbf{q})  \, \times \\
& \left\{
A_{+}^{\prime}\left(\mathbf{q}, \tau^{\prime}\right) A_{+}^{\prime}\left(\mathbf{k}-\mathbf{q}, \tau^{\prime}\right)A_{+}^{*\prime}\left(\mathbf{q}, \tau^{\prime\prime}\right)A_{+}^{*\prime}\left(\mathbf{k}-\mathbf{q},\tau^{\prime\prime}\right)+ q^2(k-q)^2 A_{+}\left(\mathbf{q}, \tau^{\prime}\right)A_{+}\left(\mathbf{k}-\mathbf{q}, \tau^{\prime}\right)
A^*_{+}\left(\mathbf{q}, \tau^{\prime\prime}\right)A^*_{+}\left(\mathbf{k}-\mathbf{q}, \tau^{\prime\prime}\right)\right.\nonumber\\
&\left.
+q(k-q)\left[A_{+}\left(\mathbf{q}, \tau^{\prime}\right)A_{+}\left(\mathbf{k}-\mathbf{q}, \tau^{\prime}\right)A_{+}^{*\prime}\left(\mathbf{q}, \tau^{\prime\prime}\right)A_{+}^{*\prime}\left(\mathbf{k}-\mathbf{q},\tau^{\prime\prime}\right)+ A_{+}^\prime\left(\mathbf{q}, \tau^{\prime}\right)A_{+}^\prime\left(\mathbf{k}-\mathbf{q}, \tau^{\prime}\right)A_{+}^{*}\left(\mathbf{q}, \tau^{\prime\prime}\right)A_{+}^{*}\left(\mathbf{k}-\mathbf{q},\tau^{\prime\prime}\right) \right] \right\}\nonumber \,. 
\end{flalign}
Assuming $A_{+}\left(\mathbf{q}, \tau'\right)=A_{+}\left(\mathbf{q}, \tau\right)T(\textbf{q},\tau,\tau')=A_{+}^\tau\left(\mathbf{q}\right)T^{\textbf{q}}_{\tau,\tau'}$ and $A^\prime_{+}\left(\mathbf{q}, \tau'\right)=A^\prime_{+}\left(\mathbf{q}, \tau\right)\bar{T}(\textbf{q},\tau,\tau')=A^{\prime_\tau}_{+}\left(\mathbf{q}\right)\bar{T}^{\textbf{q}}_{\tau,\tau'}$, where $T$ and $\bar{T}$ are real functions. We can then write
\begin{flalign}
&\left\langle{ h_{s}}(\mathbf{k}) h_{s'}\left(\mathbf{k}^{\prime}\right)\right\rangle=  \frac{2}{(2\pi)^3}\frac{H^{4} }{ m_{P}^{4}} \delta\left(\mathbf{k}+\mathbf{k}^{\prime}\right)\int d \tau^{\prime} d \tau^{\prime \prime}  \tau^{\prime 2}  \tau^{\prime\prime 2} G_{k}\left(\tau, \tau^{\prime}\right) G_{k}\left(\tau, \tau^{\prime \prime}\right)\nonumber\\
& \int d^{3}\mathbf{q}\ \epsilon_{-s}^{l}(\mathbf{k}) \epsilon^l_+(\textbf{q})\epsilon_{-s'}^{*a}(\mathbf{k}) \epsilon^{*a}_+(\textbf{q}) \epsilon_{-s}^{m}(\mathbf{k})\epsilon^m_{+}(\mathbf{k}-\mathbf{q})\epsilon_{-s'}^{*b}(\mathbf{k})\epsilon^{*b}_{+}(\mathbf{k}-\mathbf{q}) \, \times \nonumber \\
& \left\{
|A_{+}^{\prime}\left(\mathbf{q}, \tau^{\prime}\right)|^2 |A_{+}^{\prime}\left(\mathbf{k}-\mathbf{q}, \tau^{\prime}\right)|^2\bar{T}^{\textbf{q}}_{\tau',\tau''}\bar{T}^{\textbf{k-q}}_{\tau',\tau''}+q^2(k-q)^2
|A_{+}\left(\mathbf{q}, \tau^{\prime}\right)|^2|A_{+}\left(\mathbf{k}-\mathbf{q}, \tau^{\prime}\right)|^2\, T^{\textbf{q}}_{\tau',\tau''}T^{\textbf{k-q}}_{\tau',\tau''}\right.\nonumber\\
&\left.
\quad+q(k-q)\left[\bar{T}^{\textbf{q}}_{\tau',\tau''}\bar{T}^{\textbf{k-q}}_{\tau',\tau''}\,A_{+}\left(\mathbf{q}, \tau^{\prime}\right)A_{+}^{*\prime}\left(\mathbf{q}, \tau^{\prime}\right)A_{+}\left(\mathbf{k}-\mathbf{q}, \tau^{\prime}\right)A_{+}^{*\prime}\left(\mathbf{k}-\mathbf{q},\tau^{\prime}\right)\nonumber\right.\right.\\
&\left.\left.
\qquad\qquad\qquad+T^{\textbf{q}}_{\tau',\tau''}T^{\textbf{k-q}}_{\tau',\tau''}A_{+}^\prime\left(\mathbf{q}, \tau^{\prime}\right)A_{+}^{*}\left(\mathbf{q}, \tau^{\prime}\right)A_{+}^\prime\left(\mathbf{k}-\mathbf{q}, \tau^{\prime}\right)A_{+}^{*}\left(\mathbf{k}-\mathbf{q},\tau^{\prime}\right) \right] \right\}\label{Eq:4th_2point_h_modesB}.
\end{flalign}
The cross term between $A$ and $A'$ can be written as 
\begin{flalign}
A_{+}\left(\mathbf{q}, \tau^{\prime\prime}\right)A_{+}^{*\prime}\left(\mathbf{q}, \tau^{\prime\prime}\right)
=&A^{(R)}_{+}\left(\mathbf{q},\tau^{\prime\prime}\right){A_{+}^{\prime}}^{(R)}\left(\mathbf{q}, \tau^{\prime\prime}\right)+A^{(I)}_{+}\left(\mathbf{q},\tau^{\prime\prime}\right){A_{+}^{\prime}}^{(I)}\left(\mathbf{q}, \tau^{\prime\prime}\right)\nonumber\\
&+i\, \left\{A^{(I)}_{+}\left(\mathbf{q},\tau^{\prime\prime}\right){A_{+}^{\prime}}^{(R)}\left(\mathbf{q}, \tau^{\prime\prime}\right)-A^{(R)}_{+}\left(\mathbf{q},\tau^{\prime\prime}\right){A_{+}^{\prime}}^{(I)}\left(\mathbf{q}, \tau^{\prime\prime}\right)\right\}\nonumber\\
=&\mathrm{Re}[A_{+}\left(\mathbf{q}, \tau^{\prime\prime}\right)A_{+}^{\prime*}\left(\mathbf{q}, \tau^{\prime\prime}\right)]+i\, \mathrm{Im}[A_{+}\left(\mathbf{q},\tau^{\prime\prime}\right)A_{+}^{\prime*}\left(\mathbf{q}, \tau^{\prime\prime}\right)].\label{Eq:CrossAA'B}
\end{flalign}
This imaginary contribution is obtained from the normalization of the wave function giving it a constant value 1/2. With the proper renormalization one removes this vacuum contribution.
The last line in Eq. \eqref{Eq:4th_2point_h_modesB} becomes
\begin{flalign}
\bar{T}^{\textbf{q}}_{\tau',\tau''}\bar{T}^{\textbf{k-q}}_{\tau',\tau''}\mathrm{Re}[A_{+}^{\mathbf{q}}\left(\tau^{\prime}\right)A_{+}^{\prime*\,\mathbf{q}}\left(\tau^{\prime})\right] \mathrm{Re}[A_{+}^{\mathbf{k}-\mathbf{q}}\left(\tau^{\prime}\right)A_{+}^{\prime*\,\mathbf{k}-\mathbf{q}}\left(\tau^{\prime}\right)]
+T^{\textbf{q}}_{\tau',\tau''}T^{\textbf{k-q}}_{\tau',\tau''}\mathrm{Re}[A_{+}^{\prime\mathbf{q}}\left(\tau^{\prime}\right)A_{+}^{*\,\mathbf{q}}\left(\tau^{\prime}\right)]\mathrm{Re}[A_{+}^{\prime\mathbf{k}-\mathbf{q}}\left(	\tau^{\prime}\right)A_{+}^{*\,\mathbf{k}-\mathbf{q}}\left(\tau^{\prime}\right)]\,. 
\end{flalign}
To simplify the integration on $q$ we use 
\begin{equation}
\left|\epsilon_{-\lambda}^{i}\left(\mathbf{p}_{1}\right) \epsilon_{+}^{i}\left(\mathbf{p}_{2}\right)\right|^{2}=\frac{1}{4}\left(1+\lambda \frac{\mathbf{p}_{1} \cdot \mathbf{p}_{2}}{p_{1} p_{2}}\right)^{2} ,
\end{equation}
and the second line of Eq. \eqref{Eq:4th_2point_h_modesB} becomes
\begin{equation}
\int d^{3}\mathbf{q}\,\frac{1}{16}
\left(1+s \frac{\mathbf{k} \cdot \mathbf{q}}{k q }\right)
\left(1+s' \frac{\mathbf{k} \cdot \mathbf{q}}{k q }\right)
\left(1+s \frac{\mathbf{k} \cdot \mathbf{\left(k-q\right)}}{k \mathbf{\left|k-q\right|} }\right)
\left(1+s' \frac{\mathbf{k} \cdot \mathbf{\left(k-q\right)}}{k \mathbf{\left|k-q\right|} }\right) \,. 
\end{equation}
As we are numerically solving the gauge mode equations of motion until the end of inflation, with an unknown analytical solution, we will try to use the semi-analytical approach explained in section \ref{secII} to obtain these correlation functions. 
Integrating until $\epsilon_H=2$ for several modes we get the gauge mode amplitude and velocity spectrums where they are expected to be at a maximum, just before the onset of a radiation dominated universe.   
Using the transfer functions, $T$ and $\bar{T}$ we can relate both derivatives and gauge field amplitudes at a given time $\tau$ to its values at the end of inflation and to write
\begin{flalign}
&\left\langle{ h_{s}}(\mathbf{k}) h_{s'}\left(\mathbf{k}^{\prime}\right)\right\rangle=  \frac{2}{(2\pi)^3}\frac{H^{4} }{ m_{P}^{4}} \delta\left(\mathbf{k}+\mathbf{k}^{\prime}\right)\int d \tau^{\prime} d \tau^{\prime \prime}  \tau^{\prime 2}  \tau^{\prime\prime 2} G_{k}\left(\tau, \tau^{\prime}\right) G_{k}\left(\tau, \tau^{\prime \prime}\right)\nonumber\\
&	\int d^{3}\mathbf{q}\,\frac{1}{16}
\left(1+s \frac{\mathbf{k} \cdot \mathbf{q}}{k q }\right)
\left(1+s' \frac{\mathbf{k} \cdot \mathbf{q}}{k q }\right)
\left(1+s \frac{\mathbf{k} \cdot \mathbf{\left(k-q\right)}}{k \mathbf{\left|k-q\right|} }\right)
\left(1+s' \frac{\mathbf{k} \cdot \mathbf{\left(k-q\right)}}{k \mathbf{\left|k-q\right|} }\right) \, \times \\
& \left\{
|A_{+}^{\prime\,\tau_e}\left( \mathbf{q}\right)|^2 |A_{+}^{\prime\,\tau_e}\left( \mathbf{k}-\mathbf{q}\right)|^2\bar{T}^{\textbf{q}}_{\tau',\tau_e}\bar{T}^{\textbf{k-q}}_{\tau',\tau_e}\bar{T}^{\textbf{q}}_{\tau'',\tau_e}\bar{T}^{\textbf{k-q}}_{\tau'',\tau_e}+q^2(k-q)^2
|A^{\tau_e} _{+}\left(\mathbf{q}\right)|^2|A^{\tau_e}_{+}\left(\mathbf{k}-\mathbf{q}\right)|^2\,  T^{\textbf{q}}_{\tau',\tau_e}T^{\textbf{k-q}}_{\tau',\tau_e}T^{\textbf{q}}_{\tau'',\tau_e}T^{	\textbf{k-q}}_{\tau'',\tau_e}\right.\nonumber\\
&\left.\quad+q(k-q)\left(
\left|A_{+}^{\tau_e}\left(\mathbf{q}\right)A_{+}^{\prime*\,\tau_e}(\mathbf{q}) \right|
\left|A_{+}^{\tau_e}\left(\mathbf{k}-\mathbf{q}\right)A_{+}^{\prime*\,\tau_e}\left(\mathbf{k}-\mathbf{q}\right)\right| 
{T}^{\textbf{q}}_{\tau',\tau_e}{T}^{\textbf{k-q}}_{\tau',\tau_e}\bar{T}^{\textbf{q}}_{\tau'',\tau_e}\bar{T}^{	\textbf{k-q}}_{\tau'',\tau_e} \right. \right.\nonumber\\
&\left.\left. \qquad\qquad\qquad+\left|A_{+}^{\prime\tau_e}\left(\mathbf{q}\right)A_{+}^{*\,\tau_e}\left(\mathbf{q}\right)\right|
\left|A_{+}^{\prime\tau_e}\left(	\mathbf{k}-\mathbf{q}\right)A_{+}^{*\,\tau_e}\left(\mathbf{k}-\mathbf{q}\right)\right|
\bar{T}^{\textbf{q}}_{\tau',\tau_e}\bar{T}^{\textbf{k-q}}_{\tau',\tau_e}{T}^{\textbf{q}}_{\tau'',\tau_e}{T}^{\textbf{k-q}}_{\tau'',\tau_e}
\right)\right\}.\nonumber
\label{Eq:5th_2point_h_modesB}
\end{flalign}
where to simplify the notation we have written $\mathrm{Re}[A_{+}^{\tau_e}\left(\mathbf{q}\right)A_{+}^{\prime*\,\tau_e}\left(\mathbf{q})\right]=\left|A_{+}^{\tau_e}\left(\mathbf{q}\right)A_{+}^{\prime*\,\tau_e}(\mathbf{q}) \right|$.

Using the results in Eqs. \eqref{Eq:stepfunctionwithT} and \eqref{Eq:TransferfunctionT} with the functions  \eqref{Eq:A'2spectrum}, \eqref{Eq:A2spectrum} and \eqref{Eq:AA'spectrum} for the amplitudes at the end of inflation we find the result in \eqref{Eq:final_2point_h_modes}
\begin{flalign}
\left\langle h_{s}h_{s'}\right\rangle 	=&\frac{2}{(2\pi)^3}\frac{H^{4}}{{m_P}^{4}}\frac{\left(a_{e}H_{e}\right){}^{5}}{k^{8}}\text{\ensuremath{\int_{0}^{\infty}\int_{0}^{\infty}}d\ensuremath{x_{1}}d\ensuremath{x_{2}\left(\sin x_{1}-x_{1}\cos x_{1}\right)\left(\sin x_{2}-x_{2}\cos x_{2}\right)}}\nonumber
\\&\int_{0}^{\infty}d\tilde{q}\int_{0}^{2\pi}\frac{2\pi}{16}\tilde{q}^{2}d\theta(1+s\cos\theta)\left(1+s'\cos\theta\right)\left(1+s\frac{1-\frac{\tilde{q}\cos\theta}{\tilde{k}}}
{\sqrt{1+\frac{\tilde{q}^2}{\tilde{k}^2}-2\frac{\tilde{q}}{\tilde{k}}\cos\theta}}\right)\left(1+s'\frac{1-\frac{\tilde{q}\cos\theta}{\tilde{k}}}{\sqrt{1+\frac{\tilde{q}^2}{\tilde{k}^2}-2\frac{\tilde{q}}{\tilde{k}}\cos\theta}}\right)\nonumber
\\&(\tilde{q}^2(\tilde{k}-\tilde{q})^{2}\left|A_{end}\left(\tilde{q}\right)\right|^{2}\left|A_{end}\left(\tilde{k}-\tilde{q}\right)\right|^{2}T_{\tilde{q}}^{x_{1}}T_{\tilde{q}}^{x_{2}}T_{\tilde{k}-\tilde{q}}^{x_{1}}T_{\tilde{k}-\tilde{q}}^{x_{2}}\nonumber
\\&+\left|A_{end}'\left(\tilde{q}\right)\right|^{2}\left|A_{end}'\left(\tilde{k}-\tilde{q}\right)\right|^{2}\bar{T}_{\tilde{q}}^{x_{1}}\bar{T}_{\tilde{q}}^{x_{2}}\bar{T}_{\tilde{k}-\tilde{q}}^{x_{1}}\bar{T}_{\tilde{k}-\tilde{q}}^{x_{2}}\nonumber
\\&+\tilde{q}(\tilde{k}-\tilde{q})|AA_{end}'\left(\tilde{q}\right)|\left|AA_{end}'\left(\tilde{k}-\tilde{q}\right)\right|(T_{\tilde{q}}^{x_{1}}\bar{T}_{\tilde{q}}^{x_{2}}T_{\tilde{k}-\tilde{q}}^{x_{1}}\bar{T}_{\tilde{k}-\tilde{q}}^{x_{2}}+T_{\tilde{q}}^{x_{2}}\bar{T}_{\tilde{q}}^{x_{1}}T_{\tilde{k}-\tilde{q}}^{x_{2}}\bar{T}_{\tilde{k}-\tilde{q}}^{x_{1}}))\,,
\end{flalign}
where $x=-k\tau$ and $\tilde{q}=q/(a_eH_e)$.
\section{Comparison with a chaotic quartic potential}\label{appendixC}

\begin{figure}[H]
	\centering
	\includegraphics[totalheight=7cm]{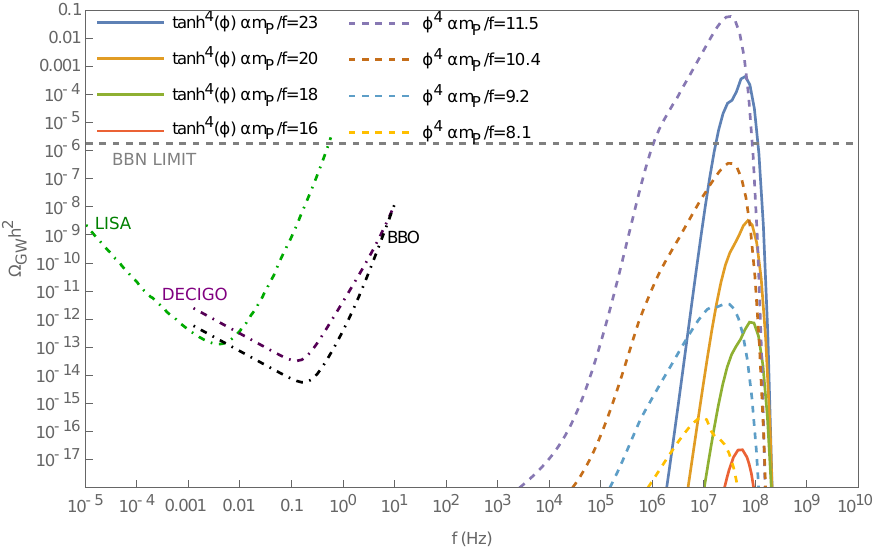}
	\caption{Comparison of the spectral energy density for the GW's produced from the tachyonic amplification estimated by the semi analytical method described with Eqs. \eqref{Eq:final_2point_h_modes} and \eqref{Eq:GWspectrum} in the cases of an $\alpha$-attractors (solid) and quartic inflationary potentials (dashed).}
	\label{fig:comp}	
\end{figure}

In this section we compare the results on the GW spectrum  with quartic inflationary potential $V(\phi)=\lambda \phi^4 /4$. 
It has been shown that an $\alpha$-attractor potential $V(\phi)=(9 \lambda/4)\tanh^4[\phi/(\sqrt{6}\, m_P)]m_P^4$ at the end of inflation and reheating coincides with the behavior from the quartic case \cite{Kallosh:2013yoa,Bastero-Gil:2020uww}. There is a natural assumption that a tachyonic production of gauge fields centered at the end of inflation would lead to similar characteristics of the spectrum of the vector modes  and the signal with gravitational waves. In Fig. \ref{fig:comp} we see which properties can be kept in the GW spectrum with the inflation model dependency in the cases with backreaction effects.

\begin{figure}[H]
	\centering
	\includegraphics[totalheight=7cm]{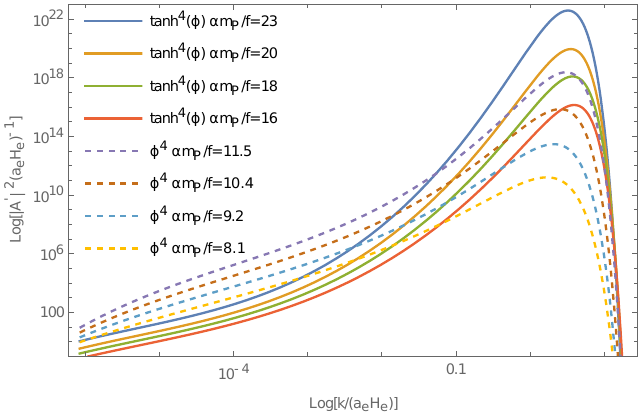}
	\caption{Comparison of the electric field $|A'(k)|^2$ spectrum generated from the tachyonic amplification in the cases of an $\alpha$-attractors (solid) and quartic inflationary potentials (dashed).}
	\label{fig:compvec}	
\end{figure}

In essence we approximately retain the peak frequencies and somehow a similar shape for the spectrum. Nevertheless, for a quartic scenario, since during the earlier stages of inflation one finds larger inflaton velocities we have broader spectrums as the lower k  modes are more amplified, see Fig. \ref{fig:compvec}. On the other hand, the vector modes in the $\alpha$-attractor case have a higher peak, revealing a larger amplification at the end of inflation.  The final GW spectrum will also depend on the time evolution of each vector mode.

We can also compare the upper bound obtained for the maximum amplitude to avoid the BBN bound. 
\begin{figure}[H]
	\centering
	
	\includegraphics[totalheight=7cm]{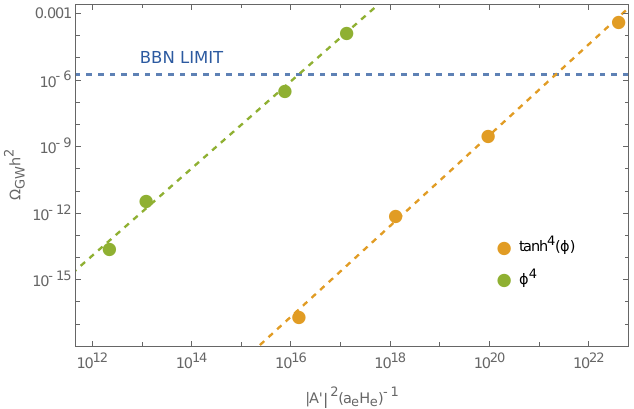}
	\caption{Comparison of the BBN bound on the amplitudes of the electric field $|A'(k)|^2$ in the cases of an $\alpha$-attractors and quartic inflationary potentials.}
	\label{fig:compBBN}	
\end{figure}

 We realize that in the $\alpha$-attractor scenario, in the no-backreaction limit, higher amplitudes for the gauge fields can be obtained without touching the BBN ceiling. This can possibly be explained due to an amplification concentrated closer to the peak, as described in Fig. \ref{fig:compvec}, and only realized very close to the end of inflation, as seen from the $\xi$ evolution in Fig. \ref{fig:xievolutionbr}. 
\cite{Bastero-Gil:2022fme}

\acknowledgments
This work has been partially supported by MICINN (PID2019-105943GB-I00/AEI/10.13039/501100011033) and ``Junta de Andaluc\'ia" grants  P18-FR-4314. ATM is supported by FCT grant SFRH/BD/144803/2019. 

%
%

\bibliography{biblio}

\end{document}